\documentclass[twocolumn,prb,aps,amsfonts]{revtex4-1}
\usepackage{latexsym}
\usepackage{dcolumn}
\usepackage[dvips]{graphicx}
\usepackage{amssymb}
\usepackage{graphicx}
\usepackage{psfrag, color}
\usepackage{verbatim}
\usepackage{amsmath}
\usepackage{epsf}

\newcommand{\vk}{{\bf k}}

\newcommand{\ve}{{\varepsilon}}

\begin{document}


\title{Transport in two-dimensional modulation doped semiconductor structures}
\author{S. Das Sarma$^1$, E. H. Hwang$^{1,2}$, S. Kodiyalam$^1$, L. N. Pfeiffer$^3$, and K. W. West$^3$}
\address{$^1$Condensed Matter Theory Center, 
Department of Physics, University of Maryland, College Park,
Maryland  20742-4111 \\
$^2$SKKU Advanced Institute of Nanotechnology and Department of
Physics, Sungkyunkwan 
University, Suwon, 440-746, Korea \\
$^3$ Department of Electrical Engineering, Princeton University, Princeton, New Jersey 08544, USA}

\date{\today}

\begin{abstract}
We develop a theory for the maximum achievable mobility in modulation-doped 2D GaAs-AlGaAs semiconductor structures by considering the momentum scattering of the 2D carriers by the remote ionized dopants which must invariably be present in order to create the 2D electron gas at the GaAs-AlGaAs interface. The minimal model, assuming first order Born scattering by random quenched remote dopant ions as the only scattering mechanism, gives a mobility much lower (by a factor of 3 or more) than that observed experimentally in many ultra high-mobility modulation-doped 2D systems, establishing convincingly that the model of uncorrelated scattering by independent random remote quenched dopant ions is often unable to describe the physical system quantitively. We theoretically establish that the consideration of spatial correlations in the remote dopant distribution can enhance the mobility by (up to) several orders of magnitudes in experimental samples. 
The precise calculation of the carrier mobility in ultra-pure modulation-doped 2D semiconductor structures thus depends crucially on the unknown spatial correlations among the dopant ions in the doping layer which may manifest sample to sample variations 
even for nominally identical sample parameters (i.e., density, well width, etc.),
depending on the details of the modulation-doping growth conditions.
\end{abstract}

\maketitle

\section{introduction}
One of the most studied physical properties in all of physics during the last 40 years has been the low-temperature electrical conductivity ($\sigma$) of two-dimensional electron gases (2DEG) in confined semiconductor structures such as Si-MOSFETs and 2D GaAs-AlGaAs heterostructures and quantum wells. Many important fundamental discoveries \cite{nayakrmp,mit,tsuiprl1982,willettprl1987,lillyprl,drag,exciton,canted,micro, wigner,spon,anyon,local} have been made in the context of studying 2DEG transport properties such as integer and fractional quantum Hall effects \cite{tsuiprl1982}, Mott variable range hoping \cite{local}, weak and strong localization \cite{local,mit}, Wigner crystallization \cite{wigner}, possible anyonic non-Abelian fractional quantum Hall effects \cite{anyon,nayakrmp}, stripe and bubble 2D interacting phases \cite{lillyprl}, 2D metal-insulator-transitions\cite{mit}, microwave-induced-resistance oscillations\cite{micro}, Coulomb drag \cite{drag}, interlayer spontaneous coherence \cite{spon}, interlayer canted antiferromagnetism \cite{canted}, 
bilayer even-denominator fractional quantum Hall effect\cite{exciton}, 
monolayer even-denominator fractional quantum Hall effect \cite{willettprl1987}, and so on. In many situations, new (and often unexpected) experimental discoveries in 2D physics become possible because of the continuous enhancement in the 2DEG mobility $\mu$ defined as $\mu \equiv \sigma/ne = e\tau/m$ where $m$ and $\tau$ are respectively the carrier effective mass and the transport relaxation time, through careful materials improvement and clever sample design. Indeed the best available current GaAs 2DEG samples routinely achieve $\mu > 10^7$ cm$^2$/Vs at low temperatures ($\alt 1K$) and not-too-low (as well as not-too-high) carrier densities ($\sim 10^{11} - 3 \times 10^{11}$ cm$^{-2}$) with the current mobility record (obtained in 2008) being $\mu \sim 4 \times 10^7$ cm$^2$/Vs (corresponding to an almost macroscopic carrier mean free path of a fraction of a mm) obtained by the Weizmann group \cite{heiblum} and the Princeton-Bell Labs group \cite{bell}. The 2D mobility (or equivalently, carrier mean free path) has thus increased by a remarkable factor of $\sim 4000$ from $\mu \sim 10^4$ cm$^2$/Vs in 1978  \cite{stormer} to $\mu \sim 4 \times 10^7$ cm$^2$/Vs in 2008  \cite{heiblum,bell} over a 30 year period in GaAs-based modulation-doped 2DEG systems (although most of this mobility enhancement occurred in the initial $10-12$ year period with the enhancement during the last 20 years being a rather modest factor of $3-4$).

This remarkable factor of 4000 enhancement in the 2D mobility has been possible because of the MBE modulation doping technique \cite{stormer} invented in 1978 which enables a spatial separation of the ionized dopants producing the 2D carriers from the 2DEG itself by locating the ionized dopant layer inside the insulator (i.e., in Al$_x$Ga$_{1-x}$As layer) some distance `$d$' (``the set-back distance'') away from the GaAs-Al$_x$Ga$_{1-x}$As interface where the 2DEG resides (on the GaAs side of the interface). This spatial separation of the ionized dopants from the 2DEG strongly suppresses the Coulomb scattering by the dopant charged impurities, leading to the strong mobility enhancement. Increasing `$d$' over the years -- the best current modulation-doped 2D samples typically have $d=70-200$ nm -- has steadily improved the mobility as Coulomb scattering by the remote dopant ions is progressively reduced by the large set-back distance. (Unfortunately `$d$' cannot be increased indefinitely since the induced carrier density in the 2DEG due to modulation doping decreases approximately as $1/d$ and thus for very large $d$, the 2D carrier density becomes far too low.) 
It is generally believed that the current high-mobility GaAs-based 2DEG is essentially immune to remote ion scattering, and the low-temperature 2D mobility in ultra-high-mobility modulation doped systems is limited entirely by the unintentional background charged impurities invariably present in the GaAs (as well as the Al$_x$Ga$_{1-x}$As) layer even in the ultra pure materials used in the modern MBE growth process.\cite{quality} The unintentional background charged impurities are typically thought to have a concentration of $\agt 10^{13}$ cm$^{-3}$ (roughly one impurity for $\sim 10^9$ lattice sites!) in the best MBE-grown materials, and a consensus has developed in the community that reducing the background impurity content in the GaAs layer would enable further improvement in the 2D mobility, eventually reaching perhaps a mobility of $\sim 100$ million cm$^2$/Vs or more in the near future.\cite{hwang2008} We note that background impurities being present (in the GaAs layer itself) are not spatially separated from the 2DEG and therefore lead to much stronger resistive scattering compared with the remote dopants for the same impurity densities. \cite{quality,hwang2008}

It is, therefore, extremely perplexing that realistic theoretical calculations of 2D mobility in modulation-doped GaAs-Al$_x$Ga$_{1-x}$As quantum wells, assuming the existence of only scattering by the random remote ionized dopants (which must always be present in order to create the 2DEG satisfying 
overall charge neutrality), give theoretical mobility values substantially lower than the actual experimentally measured maximum mobility for the same sample parameter (e.g., $d$, $n$, the remote dopant ion density $n_d$, the 2D GaAs well thickness $a$) in the best available 2DEG samples. In particular, the $T=0$ mobility of GaAs-based 2D quantum wells was calculated by the two of us \cite{hwang2008} some years ago for various values of $d$, $n$, and $n_d$. From Fig.~1(a) of Ref. \onlinecite{hwang2008} we find that for $n=n_d=3\times 10^{11}$ cm$^{-2}$ and for $d=800$ \AA \; the calculated zero-temperature mobility limited {\it only} by random remote dopant scattering is $\mu_{th} = 13 \times 10^6$ cm$^2$/Vs for a $a=300$\AA \; wide quantum well. For exactly the same sample parameters (i.e., $n = 3 \times 10^{11}$ cm$^{-2}$; $d=800$ \AA; $a=300$ \AA), by contrast, several existing modulation-doped GaAs-Al$_x$Ga$_{1-x}$As quantum well systems have $\mu_{ex} \approx 20 - 40 \times 10^6$ cm$^2$/Vs! \cite{heiblum,bell} This is remarkable because the actual theoretical mobility in a more complete calculation can only be lower than the value ($13 \times 10^6$ cm$^2$/Vs) calculated in Ref. \onlinecite{hwang2008}, i.e., $\mu_{th}$ is a theoretical limit on the maximum possible mobility since other scattering mechanisms (e.g., background unintentional charged impurities, interface roughness, alloy disorder, phonons etc.) as well as finite temperature effects can only lower the actual mobility with the expectation that $\mu_{ex} < \mu_{th}$ must always be obeyed. Thus, the paradox that $\mu_{ex} \approx 2-3\mu_{th}$ in some samples makes no sense at all and is a serious question mark on the whole theoretical frame work!

One may wonder if such a Born approximation based Boltzmann transport theory calculation of the  
2D mobility in Ref. \onlinecite{hwang2008} should be taken as quantitatively accurate, and perhaps a better transport theory will indeed produce $\mu_{th} \geq \mu_{ex}$ as it should be. This, however, should not matter in the current context of ultra-high mobilities since the effective dimensionless disorder parameter $k_F l \sim 2\times 10^5$ (where $k_F$ and $l$ are respectively the 2D Fermi wave number and the carrier transport mean free path) in this case, making the Boltzmann theory mobility calculation within the Born approximation to be essentially an exact theory (since $k_F l \sim 2 \times 10^5 \gg 1$). 

The essential puzzle is then: Why does the measured experimental mobility in some situations surpass the maximum theoretical limit (by $200 - 300$ \% and possibly more since many scattering mechanisms are being ignored) in high-mobility GaAs modulation doped systems? We emphasize that this puzzling anomaly (i.e., $\mu_{ex} > \mu_{th}$) obviously occurs in only a few ultra high mobility modulation-doped samples with the vast majority of 2D samples satisfying the expected inequality $\mu_{th} >  \mu_{ex}$.
In particular, when the 2D mobility is limited by the background disorder scattering (rather than remote dopant scattering), the agreement between theory and experiment appears to be satisfactory.

In fact, for large enough $k_F d$ ($\gg 1$), i.e., for large set-back distances (and relatively high carrier densities so that $k_F \sim \sqrt{n}$ is not very small), a simple analytical formula for the maximum possible mobility limited by remote dopant scattering
at $T=0$ in the modulation-doped 2D systems is given by
\begin{equation}
\mu = \frac{8 e (k_F d)^3}{\pi \hbar n_d},
\label{eq_1}
\end{equation}
with the corresponding mean free path given by
\begin{equation}
l = \frac{32 n^2 d^3}{n_d},
\label{eq_2}
\end{equation}
with $k_F = \sqrt{2 \pi n}$. Eqs.~(\ref{eq_1}) and (\ref{eq_2}) assume arbitrary values of the 2D carrier density $n$ and the 2D ionized dopant density $n_d$, but in any conceivable situation with the dopants producing the 2DEG, the following inequality must necessarily hold between the ionized dopant density ($n_d$) and the 2D carrier density ($n$):
\begin{equation}
n \leq n_d.
\label{eq_3}
\end{equation}
The inequality defined by Eq.~(\ref{eq_3}) simply implies that the carrier density cannot be higher than the dopant density in order to maintain the overall charge neutrality in the system.
This leads to the maximum possible theoretical 2D mobility in modulation doped structures for $n=n_d$ -- this is really the maximum possible theoretical limit since all other scattering mechanisms except for resistive scattering by the ionized dopants in the modulation doping layer are being ignored (and all these other scattering mechanisms, most notably scattering by the unintentional background charged impurities, can only bring down the 2D mobility from this limit). Putting $n= n_d$, we get the following maximum possible 2D mobility
\begin{equation}
\mu_{max} = \frac{16 \sqrt{2\pi} e d^3 \sqrt{n}}{\hbar}
\label{eq_5}
\end{equation}

Using $n=n_d=3 \times 10^{11}$ cm$^{-2}$ and $d = 800$ \AA \; in Eq.~(\ref{eq_5}) we get a maximum theoretical limit of
\begin{equation}
\mu_m = 1.7 \times 10^7 \; {\rm cm}^2/{\rm Vs}.
\end{equation}
We note that this simple analytical approximation is close to the realistic numerical result ($\mu_m = 1.33 \times 10^7$ cm$^2$/Vs) obtained in Ref.~\onlinecite{hwang2008} for the same 2D sample with $d=800$ \AA \; and $n=n_d=3\times 10^{11}$ cm$^{-2}$. This maximum analytical theoretical limit is also surprisingly low as emphasized already, compared with existing experimental results in some GaAs-AlGaAs quantum well samples where mobilities approaching $4 \times 10^7$ cm$^2$/Vs have been achieved for the same sample parameters. \cite{heiblum,bell} We note that Eq.~(\ref{eq_5}) provides a reasonably good analytical approximation for the theoretical maximum mobility limited by remote dopant scattering.
 
The conundrum that the theoretically estimated maximum mobility in modulation-doped 2D semiconductor structures often turns out to be below the actually experimentally measured mobility was already known in the 1980s.  In fact, an early paper \cite{stern1983} predicted the maximum achievable 2D mobility to be around $2 \times 10^6$ cm$^2$/Vs in the early days of modulation doping when the typical high-mobility system had mobilities below $10^5$ cm$^2$/Vs.  Improved modulation-doping and purer source materials with less background unintentional impurities soon proved this `safe' prediction of 1983 to be incorrect by a large factor since the 2D mobility reached as high as $12\times10^6$ cm$^2$/Vs already in 1989. \cite{pfeiffer1989}  Unfortunately, this puzzle has not disappeared since our predicted maximum mobility of $13\times10^6$ cm$^2$/Vs in 2008 \cite{hwang2008} is now routinely surpassed by many existing experimental samples\cite{heiblum,bell} which typically reach mobility values exceeding $20\times10^6$ cm$^2$/Vs for the same carrier density and sample details (i.e., well width and set back distance).

There are only two possible ways out of this embarrassing conundrum of the theoretical maximum mobility limit being lower than the actual experimentally measured mobilities: (1) Somehow, in spite of $k_F l \gg 1$ condition being satisfied, the Boltzmann transport theory plus Born approximation fails in providing quantitatively reliable values of the 2D carrier mobility; (2) the physical model for impurity scattering used in the Boltzmann theory is incorrect and the remote ionized dopants cannot be considered to be randomly distributed. A third possibility, where for some unknown reason $n_d \ll n$ applies, can be ruled out due to 
charge neutrality as we have already mentioned above -- in reality, $n_d \geq n$ must apply to the experimental structures. We will assume that $n_d=n$ applies throughout the current work except where $n_d$ and $n$ are individually measured experimentally.

In this article, we consider the above two options systematically, ruling out the possibility (1) by comparing theory and experiment. Then, we consider the option (2) in some depth by calculating the 2D mobility assuming various minimal models for spatial correlations in the charged impurity positions in the dopant layer, \cite{correlation,correlation_t} 
showing that very reasonable physical assumptions about spatial charged dopant correlations lead to mobility values which can be an order of magnitude (or more) larger than the theoretical limits discussed above based on the completely random impurity scattering model. It is, therefore, likely that the dopant layer in modulation doping structures has substantial correlations in charged impurity locations, and thus, without accurate direct experimental information about impurity correlations, the precise mobility limited by remote dopant scattering remains unknown. The issue of possible impurity correlations among the ionized dopants
(sometimes also referred to as `impurity clustering' in the literature) strongly enhancing 2D mobility has been discussed occasionally in the theoretical literature over the last 30 years, \cite{correlation,correlation_t,correlation_mc,kodiyalam}
but the current work is to the best of our knowledge the only comprehensive analysis of this important problem in the context of the achievement of the highest possible 2D mobilities in realistic modulation doped semiconductor structures.

The rest of this article is organized as follows: in section II we consider the theoretical mobility calculated based on the Boltzmann theory within the Born approximation, comparing the theoretical results with experimental data produced specifically to test our transport theory; in section III we provide the basic theory for transport assuming spatial correlations in the ionized impurity distribution in the doping layer; in section IV we provide calculated transport properties based on a minimal continuum impurity correlation model; in section V we develop a numerical impurity correlation model based on a discrete atomistic lattice simulation of MBE growth which includes dopant correlation effects; finally, we conclude in section VI with a discussion of our results and a summary for future outlook. 

In order to avoid any confusion, we emphasize that all our theoretical results are based on the Boltzmann transport 
theory with the scattering rate (or equivalently its inverse, the relaxation time $\tau$) calculated in the leading-order Born approximation assuming the weak scattering limit. The only scattering mechanism we consider is the scattering by remote ionized impurities in the doping layer interacting with the 2DEG through a statically screened (within RPA) Coulomb interaction. All our calculations are done at $T=0$ with full effects of a finite quasi-2D layer thickness of the GaAs quantum well quantitatively included in the calculation which tends to reduce somewhat the bare Coulomb potential from the strict 2D zero-thickness limit. Our goal is to calculate the maximum possible 2D mobility in modulation-doped structures as limited only by carrier scattering from screened Coulomb disorder arising from the remote ionized donors.

\section{Boltzmann theory for 2D transport}

We consider in this section the Boltzmann transport theory for scattering by uncorrelated random charged impurities treated in the Born approximation. \cite{andormp} The density-dependent conductivity at $T=0$ is given by
\begin{equation}
\sigma = ne^2 \tau/m,
\label{eq_7}
\end{equation}
where
\begin{eqnarray}
\frac{1}{\tau} = \frac{2\pi}{\hbar} n_d \int \frac{d^2k'}{(2\pi)^2} & & |u(|\vk-\vk'|;d)|^2 (1-\cos \theta_{\vk,\vk'}) \nonumber \\
& & \times \delta[E(\vk)-E(\vk')]|_{k=k_F},
\label{eq_8}
\end{eqnarray}
where $E(\vk) = \hbar^2 k^2/2m$ is the 2D carrier energy dispersion, and $u(q;d)$ is the Fourier transform of the screened Coulomb interaction between an electron in the 2DEG and a charged dopant impurity in the dopant layer (at a setback distance of $d$ from the 2DEG) assuming a random ensemble averaging over uncorrelated charged impurities with a density of $n_d$ per unit area.

It is important for our later discussion to emphasize precisely the approximations involved in Eq.~(\ref{eq_8}): (1) the semi-classical Drude-Boltzmann transport theory is assumed neglecting all quantum interference effects; (2) the electron-charged impurity scattering is treated in (weak-scattering) first-order Born approximation neglecting all higher-order multiple-scattering processes; (3) the charged impurities are assumed to be randomly located in the dopant layer with no spatial correlations; (4) an ensemble averaging is carried out over the random impurity locations; (5) the electron-charged impurity scattering potential is taken to be the RPA-screened Coulomb interaction with all other effects of electron-electron interaction (except for wave vector-dependent static screening by the 2DEG) neglected in the theory;
(6) consistent with most experimental ultra-high-mobility samples, we assume delta-doping, i.e., the dopant layer is taken to be a thin 2D layer (parallel to the 2DEG at a distance `$d$' away) where all the ionized dopants are localized -- if the dopant layer is a narrow quantum well with a thickness much less than $d$, as is sometimes the case, our results remain unaffected.
We mention that the approximation scheme described herein is standard and has been used extensively in the literature on transport calculations (both by us and by others) \cite{andormp}. We also mention that we will relax the third approximation (item 3 in the list of six items above) later in the paper by assuming specific spatial correlations in the random charged impurity locations in the dopant layer (in fact, this is the central point of the current work).

The calculation of the conductivity $\sigma$ (or equivalently, the mobility $\mu = \sigma/ne$) now boils down to simply evaluating the integral in Eq.~(\ref{eq_8}) which is straightforward to do once the screened Coulomb potential $u(q;d)$ is specified. The screened Coulomb disorder is given by
\begin{equation}
u(q;d) = v(q;d)/\ve(q),
\label{eq_9}
\end{equation}
where $v(q,d)$ and $\ve(q)$ are respectively the bare Coulomb disorder and the RPA dielectric function
\begin{equation}
v(q,d) = v_c(q) e^{-q d} f_i(q),
\end{equation}
and 
\begin{equation}
\ve(q) = 1- v_c(q) \Pi(q) f_e(q),
\label{eq_11}
\end{equation}
where $f_i$ and $f_e$ are respectively the form-factors associated with electron-impurity interaction and electron-electron interaction with $v_c(q)$ being simply the 2D Coulomb-interaction
\begin{equation}
v_c(q) = 2\pi e^2/\kappa q,
\end{equation}
where $\kappa = (\kappa_{GaAs} + \kappa_{AlGaAs})/2$ is the background lattice dielectric constant. The form factors $f_i$ ($\leq 1$) and $f_e$ ($\leq 1$) arise from the quasi-2D quantum mechanical nature of the quantum well confinement (i.e., from the finite thickness
of the 2D quantum well) for the 2DEG. For a strict (and idealized) 2D confinement of zero thickness $f_i=f_e=1$, but assuming a quasi-2D infinite potential quantum well confinement with a width `$a$', $f_i$ and $f_e$ are calculated to be
\begin{equation}
f_i(q)  = \frac{4}{qa} \frac{2\pi^2(1-e^{-qa/2}) + (qa)^2}{4\pi^2 +
  (qa)^2},
\label{eq_13}
\end{equation}
and
\begin{equation}
f_e(q)  = \frac{3(qa)+8\pi^2/(qa)}{(qa)^2+4\pi^2} -
\frac{32\pi^4[1-\exp(-qa)]}{(qa)^2[(qa)^2+4\pi^2]^2}.
\label{eq_14}
\end{equation}
where we assume that $\kappa_{GaAs} = \kappa_{AlGaAs}$ (which is an excellent approximation) so that image charge effects can be neglected. In Eq.~(\ref{eq_11}), $\Pi(q)$ is the $T=0$ 2D static RPA polarizability (or screening) function given by \cite{andormp}
\begin{equation}
\Pi(q) =1-\left [ 1-\left ( \frac{2k_F}{q} \right )^2 \right ]^{1/2} \theta(q-2k_F),
\label{eq_15}
\end{equation}
where $\theta(x)$ is the Heaviside step function, i.e., $\theta(x) =1$ for $x\ge 0$ and 0  for $x<0$.
Using Eqs.~(\ref{eq_7}) -- (\ref{eq_15}) and assuming the strict 2D limit (or equivalently $d \gg a$), it is straightforward to derive that the maximum mobility (i.e., the 
mobility limited only by remote charged dopants with $n_d = n$) is given by Eq.~(\ref{eq_5}), going as $\mu_{max} \approx 16 \sqrt{2\pi} e d^3 \sqrt{n}/\hbar$ in the $k_F d \gg 1$ limit.

\begin{figure}
	\centering
	\includegraphics[width=.9\columnwidth]{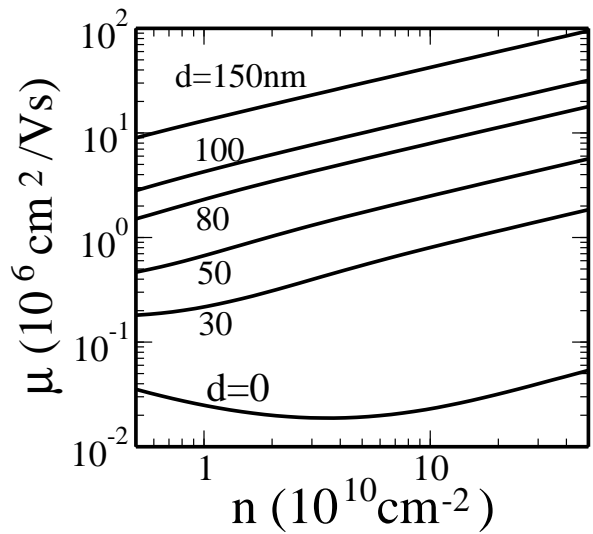}
	\caption{
Calculated mobility as a function of carrier density for different set back distances $d=0$, 30, 50, 80, 100, 150 nm. Here the impurity density $n_d=n$ is used for each plot and a quantum well width $a=300$ \AA \; is used.
}
\label{fig1}
\end{figure}

Full numerical calculations for transport properties based on Eqs.~(\ref{eq_7}) -- (\ref{eq_15}) have already been provided by us in some details in Ref. \onlinecite{hwang2008}, and therefore we refrain from further discussion of the theory except to provide in Fig.~1 one set of numerical results for the calculated (maximum possible) mobility based on Eqs.~(\ref{eq_7}) -- (\ref{eq_15}) for a 2D GaAs-AlGaAs quantum well of thickness $a=300$ \AA \; for a series of values of set back distance ($d=0$, 300, 500, 800, 1000, 1500 \AA) as a function of 2D carrier density $n$, explicitly assuming $n=n_d$ for each plot. For larger values of $k_F d$, the mobility shown in  Fig.~1 is slightly below the analytical $k_F d \gg 1$ result given in Eq.~(\ref{eq_5}), 
approaching it asymptotically for very large $k_F d$ ($k_F d \gg 1$, $d \gg a$, $k_F \gg q_{TF}$). We mention, as noted already earlier, that the calculated mobility in the very best 2D modulation-doped GaAs structures \cite{heiblum,bell}
routinely falls well below the experimentally measured mobility for large values of $d$
although the vast majority of 2D GaAs samples obviously has mobility lower that the theoretical predictions in Fig.~1 which, in principle, should be an absolute upper limit on the maximum possible mobility for given values of $n$ and $d$ for all samples.

To establish that the Boltzmann theory developed in Eqs.~(\ref{eq_7}) -- (\ref{eq_15}) is quantitatively reliable for high-mobility ($k_Fl \gg 1$) samples
(at least in some situations) we show in Fig.~2 a comparison between the theory and experimental data. These experimental results are specifically obtained in order to assess the quantitative validity of the Boltzmann transport theory in a 2D HIGFET (heterostructure insulated gate field effect transistor) structure \cite{lilly} where the 2D carrier density can be tuned by changing the voltage on an external gate (thus, the HIGFET is the GaAs-AlGaAs version of the Si-SiO$_2$ MOSFET -- metal oxide semiconductor field effect transistor). \cite{andormp}  Before describing the results in any details (see below) we emphasize that Fig.~2 directly establishes the quantitative accuracy and the qualitative validity of the Boltzmann transport theory for 2D transport calculations through the comparison of the density dependent mobility in 2D GaAs between theory and experiment without any adjustable parameters.

\begin{figure}
	\centering
	\includegraphics[width=.9\columnwidth]{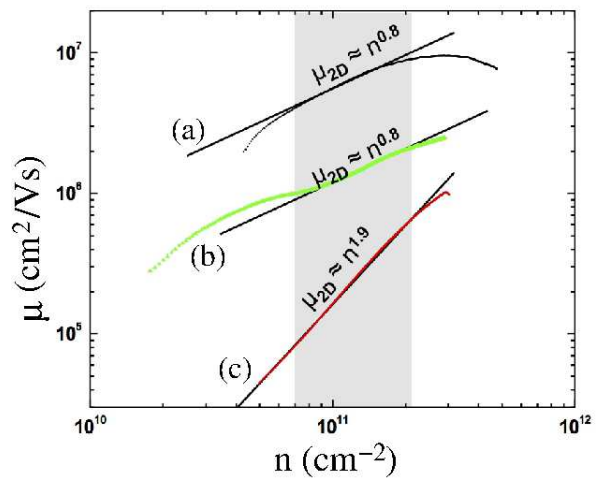}
	\caption{
The measured 2D mobility compared with the theoretical Boltzmann calculation (solid lines).
In (a) and (b) $n_i=10^{14}$ cm$^{-3}$ and $n_i=5\times 10^{14}$ cm$^{-3}$ of C atoms, respectively, are introduced in $a=300$ \AA \;wide  2D GaAs channel.
In (c) $n_i=10^{17}$ cm$^{-3}$ of C atoms are introduced in the AlGaAs barrier over a 10 nm wide strip located at a set back distance of 10 nm from the GaAs-AlGaAs interface.
The experiments are carried out at $T=300$ mK.
}
\label{fig2}
\end{figure}

The key differences between HIGFETs \cite{lilly} and  modulation-doped structures (of interest in the current work) are the following: (1) there are no remote dopants in HIGFETs in contrast to modulation-doped structures; (2) the carrier density can be tuned in HIGFETs by the external gate (similar to MOSFETs) which supplies the 2D carriers in contrast to modulation doped structures where the carriers are induced by the remote dopants in the doping layer; (3) the 2D mobility in high-quality HIGFETs is limited 
entirely by background unintentional charged impurity scattering and the whole issue of spatial correlations among donor charged impurity locations (in the dopant layer of the modulation doped system) is moot in HIGFETs since the unintentional background charged impurities are most likely distributed randomly spatially in an uncorrelated manner; (4) the  random uncorrelated spatial distribution of background impurities (in contrast to possible spatially correlated charged impurities in the dopant layer) can be treated using our Boltzmann transport theory within the Born approximation as long as the mean free path is sufficiently long so that the higher order impurity scattering effects (beyond the leading-order Born approximation) are neglected; (5) these differences between HIGFETs and modulation-doped high mobility structures indicate that while the mobility in the former is limited entirely by scattering from unintentional background charged impurities (which are likely to be distributed in spatially uncorrelated random locations), the mobility in the latter is limited by the combined Coulomb scattering from both unintentional background charged impurities and remote charged dopants.

In Fig.~2 we show that the measured 2D mobility compared with the theoretical Boltzmann calculation (using the Born approximation and random ensemble averaging over uncorrelated background impurity positions). Results for three distinct experimental measurements are shown where we have taken care to experimentally introduce calibrated background disorder into the HIGFET sample. First, we introduce $10^{14}$ cm$^{-3}$ of Carbon atoms in the 2D GaAs channel itself; second, we introduce $5\times 10^{14}$ cm$^{-3}$ of C atoms in the 2D GaAs channel, and 
finally, we introduce $10^{17}$ cm$^{-3}$ of C atoms in the AlGaAs barrier over a 10 nm wide strip located at a set back distance of 10 nm from the GaAs-AlGaAs interface.

We note that the residual background impurity concentration (i.e., before the controlled incorporation of additional impurities) in our HIGFET sample is, of course, unknown by definition (since these are unintentional impurities), but is thought to be rather small ($\leq 10^{13}$ cm$^{-3}$). What is clear from Fig.~2 is that the theory, based on the Boltzmann theory and Born approximation, agrees {\it quantitatively} with the experimental data in the sense that (1) increasing the background impurity density by a factor of 5 (from $10^{14}$ cm$^{-3}$ to $5\times 10^{14}$ cm$^{-3}$) exactly decreases the mobility by a factor of 5, and (2) the calculated carrier density dependence of mobility ($\mu \sim n^{0.8}$ or $\mu \sim n^{1.9}$) agrees very well with the experimental measurement over a broad range of intermediate density (where the theory is particularly well-valid) in all the experimentally investigated situations including the lowest mobility case, where $\mu \sim n^{1.9}$, involving a larger density ($=10^{17}$ cm$^{-3}$) of added charge impurities in the barrier. \cite{density} We note that the theory becomes progressively inaccurate at both high and low densities (and hence the systematic disagreement between theory and experiment at high and low carrier density regimes in Fig.~2) since other physical mechanisms (beyond screened Coulomb disorder scattering arising from random background charged impurities) come into play in these regimes. In the high-density regime, the self-consistent electric field generated by the combined 2D electrons themselves becomes important which pushes the electrons closer to the GaAs-AlGaAs interface as the 2D confinement becomes tighter. This tighter confinement at higher density leads to the carrier scattering from the interface roughness and from the alloy disorder in the barrier which become quantitatively important \cite{andormp}, eventually reducing the mobility with increasing density at high enough carrier density. Although it is straightforward to include this short-range disorder effect in the theory, it is beyond the scope of the current work where we are only considering effects of Coulomb disorder. At low carrier densities, where screening is weak, the random charged impurities could lead to strong density inhomogeneity (``puddle") in the sample (and associated carrier localization and metal-insulator transition). \cite{fluctuation} These low-density effects of incipient carrier localization and percolation through inhomogeneous density fluctuations are not included in the Boltzmann transport theory with Coulomb disorder treated in the static RPA screening approximation. Thus, the measured mobility eventually falls below the calculated mobility at low and high carrier densities, but the agreement between theory and experiment is excellent in the intermediate density regime where the theory is particularly well-valid.

Before concluding the discussion of the experimental results (and their comparison with the Boltzmann theory with Coulomb disorder scattering) in HIGFETs, we mention that the theoretical results for the lowest mobility sample in Fig.~2, where $10^{17}$ cm$^{-3}$ of C atoms are inserted into the AlGaAs barrier region in a 10 nm strip located 10 nm away from the interface, are easily obtained through a minor modification of Eq.~(\ref{eq_8}) where $n_d |u(\vk-\vk')|^2$ is replaced by 
\begin{equation}
\int_{-\infty}^{\infty} dz N_i(z) |u(\vk-\vk',z)|^2,
\end{equation}
inside the $\vk'$ integration with $N_i(z)$ being the 3D impurity density in the system. For the lowest mobility sample of Fig.~2, we have $N_i(z) = N_i = 10^{17}$ cm$^{-3}$ only for the strip $z=200$ \AA $-300$ \AA \; where $z$ is now measured from the GaAs-AlGaAs interface. The theoretical calculations for the HIGFET use realistic HIGFET quasi-2D confinement effect with the form-factors $f_i$ and $f_c$ (see Eqs.~(\ref{eq_13}) and (\ref{eq_14})) being calculated variationally for the appropriate triangular 2D confinement in HIGFETs in contrast to the square well confinement in modulation-doped quantum wells. Since these details are well-established in the literature, we refrain from giving the relevant theoretical formula instead referring to the existing literature. \cite{andormp}

We emphasize that the theory describes well both the quantitative aspects of the measured mobility as well as the qualitative aspect namely the fact, that the mobility limited by channel impurities (barrier impurities) manifests completely different density scaling behavior with $\mu \sim n^{0.8}$ for channel impurities and $\mu \sim^{1.9}$ for barrier impurities. This density scaling behavior of 2D transport has recently been discussed in details by us elsewhere \cite{density} and we refrain from further discussion of this point except to note that the power law exponent $\alpha$ with $\mu \sim n^{\alpha}$ can be used experimentally to distinguish whether the mobility in a particular sample in a given density range is being dominated by nearby channel (or far away barrier) impurities by measuring whether $\alpha < 1$ or $\alpha > 1$ respectively.

\section{general theory for impurity correlations}

Having established in the last section the basic validity of the Boltzmann transport theory (with impurity scattering treated in Born approximation) for spatially uncorrelated random impurity scattering (both in the 2D GaAs channel and in the AlGaAs barrier region) by directly comparing experiment and theory, we now face the conundrum produced by some of the highest mobility modulation doped quantum well structures reported in the literature, where the measured mobility exceeds the theoretically allowed maximum mobility calculated within the Boltzmann theory assuming spatially uncorrelated random scattering by the remote charged dopants (without any other scattering).  Making the situation even more mysterious is the fact that the mobility in these extreme high mobility ($\mu > 2 \times 10^7$ cm$^2$/Vs for $n \sim 3\times 10^{11}$ cm$^{-2}$) modulation-doped structures is known to be limited {\it not} by the remote dopant scattering, but by the scattering from unintentional background charged impurities in the GaAs layer itself since the density dependence of the mobility, $\mu \sim n^{\alpha}$, follows a power law with the scaling exponent $\alpha \sim 0.7 - 0.8$ indicating the strong dominance of background scattering over remote scattering. \cite{density} If remote dopant scattering (i.e., $k_F d \gg 1$) is the dominant scattering mechanism, then the density  
power law exponent $\alpha$ should be larger, $\alpha \agt 1.5$, which is never seen in these extreme high-mobility modulation-doped structures. On the other hand, as emphasized in depth in the Introduction (and as can be seen in Fig.~1), the calculated mobility for $n=3\times 10^{11}$ cm$^{-2}$ due to remote dopant scattering (with $n_d = n$) gives $\mu_{th} \approx 1.3 \times 10^7$ cm$^2$/Vs which is less that the measured mobility, $\mu \approx 2-4 \times 10^7$ cm$^2$/Vs. Obviously, something is missing in the theory!

We now consider spatial correlations among the locations of the charged dopants as the main physical mechanism enhancing the 2D mobility limited by remote scattering (compared with the theoretical results shown in Fig.~1 which are obtained by ensemble averaging over random spatially uncorrelated positions of the ionized donor impurities in the remote doping layer). Formally, it is  
straightforward to modify the Boltzmann transport theory including spatial correlations among the impurity locations within the leading-order Born approximation. In the presence of inter-impurity spatial correlations, \cite{correlation,correlation_t} Eq.~(\ref{eq_8}) for the transport scattering rate changes to
\begin{eqnarray}
\frac{1}{\tau} = \frac{2\pi}{\hbar} & & \int \frac{d^2k'}{(2\pi)^2}  s_d({\bf k-k'})|u({\bf k-k'};d)|^2
  \nonumber \\
& & \times (1-\cos \theta_{\vk\vk'})  \delta(E(\vk)-E(\vk'))|_{k=k_F},
\label{eq_16}
\end{eqnarray}
where the remote dopant charged impurity density $n_d$ in Eq.~(\ref{eq_8}) has been replaced by the spatially correlated impurity structure factor $s_d({\bf q})$ defined by
\begin{equation}
s_d({\bf q}) = \left | \sum_{i=1}^{N_i} e^{-i {\bf q \cdot R}_i} \right |^2,
\label{eq_17}
\end{equation}
where ${\bf R}_i$ is the 2D location of the $i$-th
ionized dopant in the doping layer with $i$ denoting the individual ionized dopant site with the sum ranging over all $N_i$ of the ionized dopants (i.e., $N_i$ is the total number of ionized dopants in the modulation doping layer whereas $n_d$ is their 2D areal density).

An easy way to derive Eqs.~(\ref{eq_16}) and (\ref{eq_17}) is to use the Fermi's golden rule and the weak-scattering Born approximation to calculate the transport relaxation rate from the total scattering potential created by the ionized dopants
\begin{eqnarray}
\frac{1}{\tau} = \frac{2\pi}{\hbar} \int \frac{d^2k'}{(2\pi)^2} & & \left | \tilde{u}(\vk-\vk';d) \right |^2 ( 1- \cos \theta_{\vk\vk'}) \nonumber \\
& & \times \delta[E(\vk) - E(\vk')]|_{k=kF}
\label{eq_18}
\end{eqnarray}
where $\tilde{u}({\bf q};d)$ is the Fourier transform of the 
electrostatic potential created by the ionized dopants
\begin{equation}
\tilde{u}({\bf q};d) = \sum_{i=1}^{N_i} \int d^2 r u({\bf r-R}_i,d) e^{-i {\bf q \cdot (r-R}_i)} e^{-i {\bf q \cdot (r-R}_i)},
\label{eq_19}
\end{equation}
where (${\bf R}_i,d$) is the position of the $i$-th ionized dopant with ${\bf R}_i$ being the 2D position vector and $d$ the set-back distance of the dopant layer. The potential $u({\bf r-R}_i,d)$ is the screened Coulomb potential arising from the $i$-th charged donor in the dopant layer at the 3D position vector $(x,y) \equiv {\bf R}_i$ and $z=d$ at the electron position {\bf r} in the 2DEG. Rewriting Eq.~(\ref{eq_19}) as 
\begin{equation}
\tilde{u}({\bf q};d) = \sum_{i=1}^{N_i} e^{-i{\bf q\cdot R}_i} \int d^2 r' u({\bf r'},d) e^{-i {\bf q \cdot r'}},
\label{eq_20}
\end{equation}
where ${\bf r' = r-R}_i$ has been used, we get immediately 
\begin{equation}
\tilde{u}({\bf q};d) = u(q,d) \left ( \sum_{i=1}^{N_i} e^{-i{\bf q\cdot R}_i} \right ).
\label{eq_21}
\end{equation}
Substituting Eq.~(\ref{eq_21}) into Eq.~(\ref{eq_18}) leads to Eq.~({\ref{eq_16}).
The basic idea of including the structure factor in the scattering formula [Eq.~(\ref{eq_16}) above] is that the scattering cross-section is not determined by a sum over individual scattering by each independent point scatterer, but instead by the scattering from the total potential created by all the scattering centers together.  Since there could be considerable interference among the scattering events from the individual charged impurities, particularly when there are spatial correlations among their locations, the correlated theory as defined by Eq.~(\ref{eq_16}) would in general have suppressed scattering compared with the simple ensemble averaged random impurity formula summing up all the individual scattering events.  Thus, spatial correlations would make $s_d<n_d$, leading in general to higher mobility.

If the impurities are completely uncorrelated, then $s_d({\bf q}) = n_d$, which leads to Eq.~(\ref{eq_18}) for scattering by random spatially uncorrelated charged impurities. In general, however, spatial correlations lead to $s_d < n_d$ because the presence of correlations leads to some interference between scattering from different
impurity locations. Thus, spatial impurity correlations always reduce the scattering rate, and hence enhances $\tau$, thus leading to increased conductivity ($\propto \tau$) and mobility ($\propto \tau$). The magnitude of the actual mobility enhancement due to impurity correlations in a specific sample depends entirely on the level of inter-impurity correlations in the sample itself. A very special case is that of the ionized donors in the dopant layer being spatially arranged in a perfect 2D periodic structure (e.g., a 2D square lattice or a honeycomb structure) where the only possible scattering by the ionized impurities is Bragg scattering, and therefore, the scattering rate $\tau^{-1}$ vanishes identically (at $T=0$), leading to infinite mobility or conductivity! Thus, the actual 2D mobility is limited by remote ion scattering from below by the results shown in Fig.~1 where the remote ions are assumed to be randomly located in the modulation doping layer with no spatial correlations and is bounded from above by infinite mobility where the ionized dopants form a perfect 2D crystal (so that the random location of one impurity determines the spatial positions of all the other impurities).

The actual 2D mobility lying between its random impurity value (in Fig.~1) and infinity is going to depend on the delicate details of the spatial correlations among the charged impurities in specific samples. Obviously, the precise impurity spatial correlations could never be known experimentally (or otherwise). In the next two sections, we will discuss various approximations for incorporating impurity spatial correlations in the transport theory.
It is, however, clear that any spatial correlations among the ionized donors in the modulation doping layer will enhance the 2D mobility above its value shown in Fig.~1
bringing theory and experiment closer.

\section{continuum model of impurity correlations}

In the continuum model, the impurity structure factor $s_d({\bf q})$ of Eq.~(\ref{eq_17}) can be rewritten in terms of the real space pair correlation function $g_d({\bf r})$ defining the spatial correlations among the charged donors
\begin{equation}
s_d({\bf q}) = n_d \left [ 1 +  n_d\int d^2r e^{i{\bf q \cdot r}} \left \{g_d({\bf r}) -1 \right \} \right ].
\end{equation}
The transport calculation in the presence of impurity spatial correlations now boils down to making suitable approximations for the charged dopant pair correlation function $g_d({\bf r})$.

\begin{figure}
	\centering
	\includegraphics[width=1.\columnwidth]{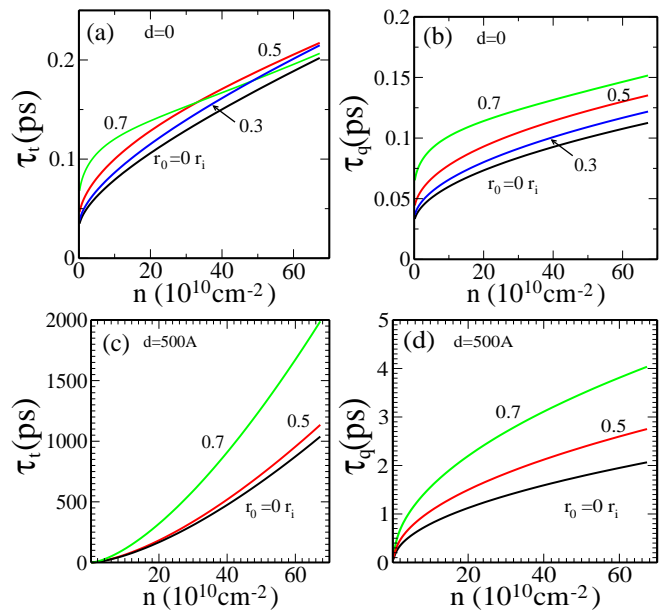}
	\caption{
Calculated (a) transport scattering time $\tau_t$ and (b) quantum scattering time $\tau_q$ as a function carrier density for different correlation length $r_0 = 0,$ 0.3, 0.5, 0.7$r_i$, where $r_i = 1/(\pi n_d)^{1/2}$. The results are calculated with $d=0$ and a fixed impurity density $n_d = 10^{11} cm^{-2}$. 
(c) and (d) show the same as (a) and (b), respectively, but with a set back distance of $d=500$ \AA.
}
\label{fig3}
\end{figure}

\begin{figure}
	\centering
	\includegraphics[width=1.\columnwidth]{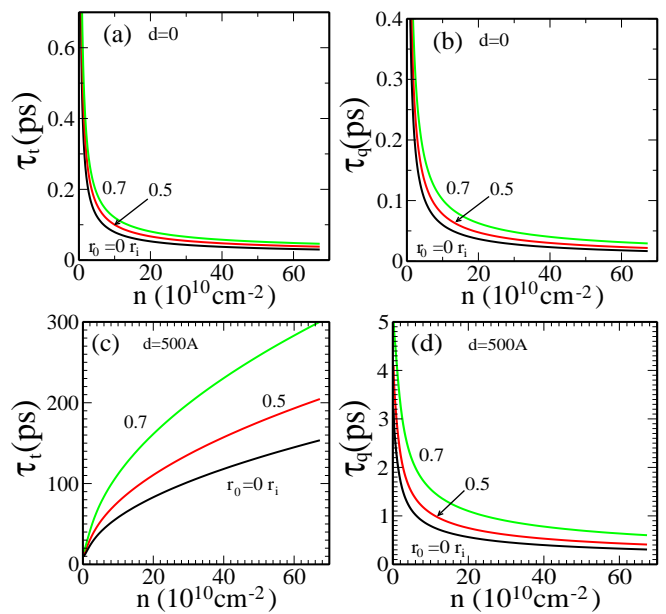}
	\caption{
Calculated (a) transport scattering time $\tau_t$ and (b) quantum scattering  time $\tau_q$ as a function carrier density for different correlation length $r_0 = 0,$ 0.5, 0.7$r_i$. The results are calculated with $d=0$ and $n_d = n$. 
(c) and (d) show the same as (a) and (b), respectively, but with $d=500$ \AA.
}
\label{fig4}
\end{figure}

The simplest possible approximation one can make for the pair correlation function $g_d({\bf r})$ is the following
\begin{eqnarray}
g_d({\bf r}) & = & 0, \;\;\; |{\bf r}| \le r_0 \nonumber \\
                  & = & 1, \;\;\; |{\bf r}| > r_0
\label{eq_24}
\end{eqnarray}
with $r_0 < r_i = (\pi n_d)^{-1/2}$ by definition. The approximation defined by Eq.~(\ref{eq_24}) asserts that any two charged donors in the dopant layer simply cannot be separated by a distance closer than a minimum distance $r_0$ -- for example, two ionized dopants are not allowed to be on top of each other (i.e. $r_0 = 0$) which is permissible for an uncorrelated random distribution of the donor ions. This minimal spatial correlation model defined by Eq.~(\ref{eq_24}) leads to 
\begin{equation}
s_d({\bf q}) = n_d \left [ 1- 2\pi n_d \frac{r_0}{q} J_1 (q r_0) \right ],
\label{eq_25}
\end{equation}
where $J_1(x)$ is the Bessel function of the first kind. The simplicity of the continuum model defined
by Eq.~(\ref{eq_24}) enables us to characterize the impurity spatial correlation effects by a single length parameter $r_0$, the minimum inter-impurity separation allowed among the 
spatially correlated ionized donors.
Note that in Ref.~[\onlinecite{correlation_t}] the scattering time was calculated by considering
the correlation effect of charged impurities, taking into account both fluctuations in the distribution of the impurity sites (impurity configuration) and the charge state (negative os positive) of a given impurity.
In addition, the Thomas-Fermi screening approximation was used 
to treat the charged impurity scattering potential.
It was shown \cite{correlation_t} that an enhancement in the mobility was observed when the impurity charge state fluctuations were large. However, the mobility enhancement arising from the spatial correlations among the charged donors  was not discussed. 
In our model we consider one type of charged impurities (positively charged dopants) and therefore only the spatial correlations among the charged impurities in contrast to Ref.~[\onlinecite{correlation_t}].

Inserting Eq.~(\ref{eq_25}) in Eq.~(\ref{eq_16}), and carrying out the momentum integration in the $k_Fd \gg 1$ regime of interest for modulation-doped quantum wells, we get (for $n_d r_0^2 \ll 1$)
\begin{equation}
\tau^{-1} \propto \frac{n_d}{(k_Fd)^3} \left [ 1 - \pi n_d r_0^2 \right ] + O(k_Fd)^{-5}.
\label{eq_26}
\end{equation}
The spatial impurity correlation induced suppression of the transport scattering rate by the correlation factor of ($1-\pi n_d r_0^2$) leads to the following relationship between the theoretical mobility calculated using uncorrelated random impurity model ($\mu_r$) as obtained from Eq.~(\ref{eq_8}) (and given in Fig.~1) 
and that calculated ($\mu_c$) using spatially correlated impurities to be
\begin{equation}
\frac{\mu_c}{\mu_r} \approx \frac{1}{1-\pi n_d r_0^2}.
\label{eq_27}
\end{equation}
For $r_0 \ll 1/\sqrt{\pi n_d}$, $\mu_c \approx \mu_r ( 1+ \pi n_d r_0^2)$, but for $r_0 \alt 1/\sqrt{\pi n_d}$, when impurity spatial correlations are strong $\mu_c \gg \mu_r$. We believe that the ultrahigh mobility modulation-doped structures typically are in this latter situation with $\pi n_d r_0^2 \alt 1$. We note that the specific form of Eq.~(\ref{eq_27}) applies only in the $r_o \ll (\pi n_d)^{-1/2}$ limit.

\begin{figure}
	\centering
	\includegraphics[width=1.\columnwidth]{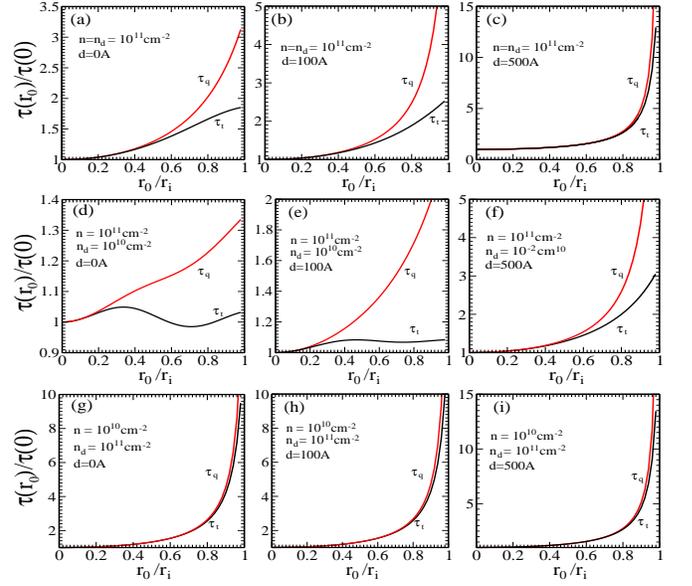}
	\caption{
(a), (b), and (c) show scaled scattering times [$\tau_{t,q}(r_0)/\tau_{t,q}(0)$] as a function of correlation length $r_0/r_i$ for (a) $d=0$, (b) $d=100$ \AA, and (c) $d=500$ \AA, and for $n=n_d=10^{11}cm^{-2}$. Here $\tau_{t,q}(0) = \tau_{t,q}(r_0=0)$. 
(d), (e), and (f) show $\tau_{t,q}(r_0)/\tau_{t,q}(0)$  for (d) $d=0$, (e) $d=100$ \AA, and (f) $d=500$ \AA \; with $n=10^{11}cm^{-2}$ and $n_d=10^{10}cm^{-2}$, i.e., $n>n_d$.
(g), (h), and (i) show $\tau_{t,q}(r_0)/\tau_{t,q}(0)$ for (g) $d=0$, (h) $d=100$ \AA, and (i) $d=500$ \AA with $n=10^{10} cm^{-2}$ and $n_d=10^{11}cm^{-2}$, i.e. $n<n_d$.	
}
\label{fig5}
\end{figure}

\begin{figure}
	\centering
	\includegraphics[width=1.\columnwidth]{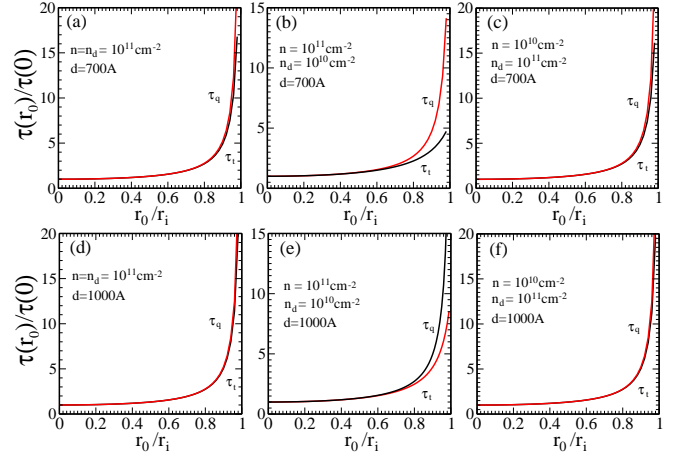}
	\caption{
(a), (b), and (c) show $\tau_{t,q}(r_0)/\tau_{t,q}(0)$ as a function of correlation length $r_0/r_i$ for a fixed set back distance $d=700$ \AA \; and
for (a) $n=n_d$, (b) $n>n_d$, and (c) $n<n_d$.
(d), (e), and (g) show $\tau_{t,q}(r_0)/\tau_{t,q}(0)$ for $d=1000$ \AA \; and 
for (d) $n=n_d$, (e) $n>n_d$, and (f) $n<n_d$.
}
\label{fig6}
\end{figure}

In Figs.~3 -- 8 we show our full numerical results from the calculated scattering time $\tau$ limited only by remote scattering in the presence of ionized dopant spatial correlation effects characterized by the dimensionless correlation parameter $r_0/r_i \equiv r_0/(\pi n_d)^{-1/2}$ (i.e., $r_i=(\pi n_d)^{-1/2}$) with $0 \leq r_0/r_i \leq 1$ with the lower ($r_i=0$) and the upper ($r_i = r_0$) bound indicating 
the standard uncorrelated random impurity distribution and the (hypothetical) perfectly periodic impurity distribution respectively. Since there are four distinct length variables, $\tau=\tau(n,n_d,d,r_0)$, characterizing the 2D transport properties (even at $T=0$ and neglecting all scattering processes other than scattering by charged impurities of 2D density $n_d$ located at a distance $d$), there is a vast amount of different results our numerics can produce. We, therefore, only provide some snapshots of representative results in order to establish the importance of impurity correlations in determining the 2D transport behavior.

In addition to showing the transport scattering time ($\tau_t$ in Figs.~3--8) discussed above in Eqs.~(\ref{eq_1})--(\ref{eq_27}) defining the 2D conductivity and mobility
$\sigma = ne^2 \tau_t/m$ and $\mu = e \tau_t/m$ (with $\sigma = n e \mu$ as usual), we also show the so-called ``quantum" (or ``single particle") scattering time $\tau_q$, which is an independent relaxation time defining the single-particle level-broadening $\gamma_q$ (e.g., the Dingle temperature $T_D = \Gamma_q/k_B$) through $\Gamma_q = \hbar/2\tau_q$ in the system, in our results. The quantum scattering time $\tau_q$ is simply obtained by replacing the factor $(1-\cos \theta)$ in the equations above for $\tau$ (e.g., Eqs.~(\ref{eq_8}), (\ref{eq_16}), (\ref{eq_18})) by unity (i.e. without the ($1-\cos\theta$) factor inside the integral over 2D wave vector on the right hand side) -- thus, $\tau = \tau_t$ ($\tau_q$) with the ($1-\cos\theta$) factor being present (absent) inside the wave vector integral for $\tau$. The presence (absence) of the ($1-\cos\theta$) factor implies the absence (presence) of forward scattering contributions to the scattering rate $\tau^{-1}$ respectively, indicating
the irrelevance (relevance) of carrier scattering in the forward direction (i.e. $\theta=0$) contributions to $\tau_t^{-1}$ ($\tau_q^{-1}$). It is obvious that $\tau_t \geq \tau_q$ with the equality holding only when the scattering potential $u({\bf q})$ is purely symmetric (i.e., pure $s$-wave) being independent of {\bf q}, which happens only for zero-range isotropic scatterers. It is clear (and well-known) that for remote scattering by ionized dopants in the modulation doping layer, $\tau_t \gg \tau_q$ since essentially all the scattering is in the forward direction, \cite{dassarmaPRB1985} and it is easy to show that in the limit $k_F d \gg 1$, $\tau_t/\tau_q \sim (k_F d)^2$. On the other hand, for unintentional background impurity scattering, $k_F d \ll 1$ and $\tau_t \sim \tau_q$ essentially
since screening makes the Coulomb disorder arising from the background charged impurities to be effectively short-ranged in nature whereas screening is relatively ineffective when the charged impurities are very far away and most of the scattering is basically small-angle forward scattering \cite{quality}.

\begin{figure}
	\centering
	\includegraphics[width=1.\columnwidth]{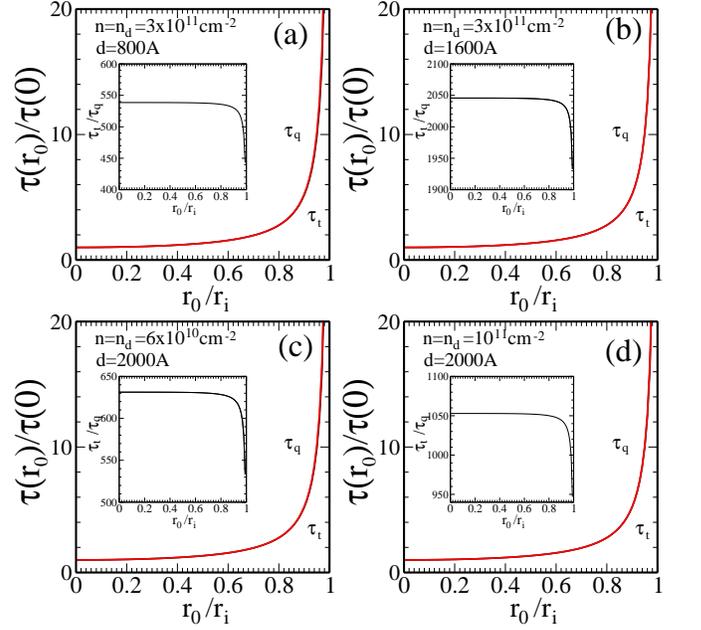}
	\caption{
(a) and (b) show scattering times $\tau_{t,q}(r_0)/\tau_{t,q}(0)$ as a function of correlation length $r_0/r_i$ for $n=n_d=3\times10^{11} cm^{-2}$ and 
for different set back distances (a) $d=800$ \AA \; and (b) $d=1600$ \AA.
(c) and (d) show 
$\tau_{t,q}(r_0)/\tau_{t,q}(0)$ for a fixed set back distance $d=2000$ \AA, and (c) $n=n_d=6\times10^{10} cm^{-2}$ and (d) $n=n_d=10^{11} cm^{-2}$.
Insets show the ratio of scattering times, $\tau_t/\tau_q$, as a function of correlation length.
}
\label{fig7}
\end{figure}

The full numerical results in Figs.~3--8 clearly show the great importance of spatial correlations, particularly for $r_0/r_i \agt 0.8$ (see Fig.~8). Although the spatial correlation effects are fairly modest for $r_0/r_i < 0.8$, the mobility (i.e., $\tau_t$) could still easily be enhanced by factors of $2-3$ even for $r_0/r_i \approx 0.7$ compared with the uncorrelated random scattering situation (Fig.~1) of $r_0=0$. For larger $r_0/r_i$ ($>0.7$), however, both $\tau_t$ and $\tau_q$ are strongly enhanced compared with their random uncorrelated impurity scattering values. In addition, it is apparent, particularly from Figs.~5--8, that a very small change in $r_0/r_i$ (for $r_0/r_i >0.7$) could have a large effect on $\tau_t$ and $\tau_q$ (and hence, on the mobility and the level broadening).

\begin{figure}
	\centering
	\includegraphics[width=1.\columnwidth]{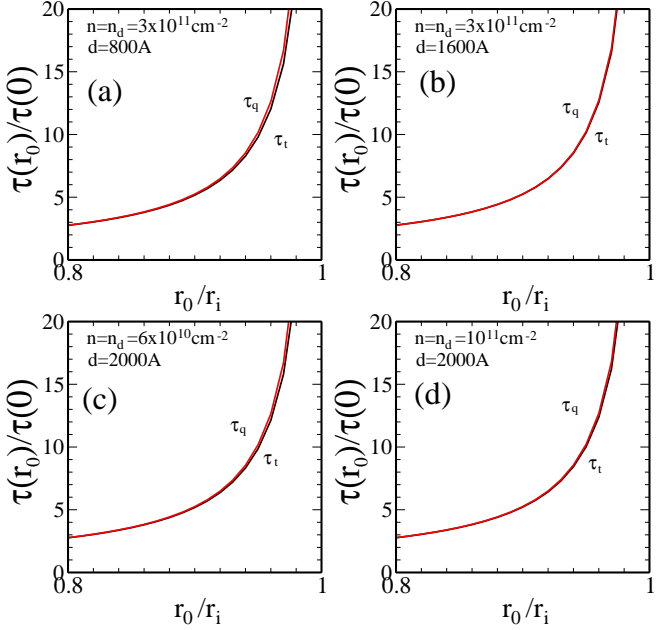}
	\caption{
The same results of Fig.~\ref{fig7} shown only for the range of $0.8 < r_0/r_i < 1$.
}
\label{fig8}
\end{figure}

This extreme sensitivity of modulation-doped samples
to spatial correlations among the dopant ions in ultrahigh mobility modulation doped structures could, in principle, be a possible reason for the unexpected (and unexplained) observed large variations (by upto a factor of 2) in the low-temperature mobility of 2D modulation-doped GaAs-AlGaAs systems (in the highest mobility structures) in various samples of identical carrier density cut 
from the same large wafer. \cite{pw} 
It is, in principle, possible for the value of the spatial correlation parameter $r_0$ to vary slightly over the large area wafer, and our results in Figs. 5 -- 8 show that even a small variation in $r_0$ could lead to large variations in the mobility in the highest quality 2D samples. A similar explanation could also apply (at least as a matter of principle) to the observed variation in the mobility of the same sample (again, only in ultrahigh mobility samples with $\mu > 10^7$ cm$^2$/Vs) as the sample is thermally cycled between room temperatures and cryogenic temperatures. Warming the sample to the room temperature ($\sim 300$ K) and then cooling it back down to $\sim 50$ mK could certainly slightly modify the spatial correlations among the ionized donor impurities in the dopant layer with a concomitant slight change in the value of the correlation parameter $r_0$ leading to a change (again by upto a factor of 2 or so) in the 2D mobility. Of course, whether the underlying reason for the observed variation in the sample mobility under thermal cycling (or in the sample to sample variation in the mobility over a large wafer) is some small variation in $r_0$ as proposed here can only be ascertained by future controlled experiments focussing on this issue. We, however, raise this possibility as a matter of principle.

One aspect of the results presented in Figs. 3 -- 8 needs special emphasis. The quantum scattering time $\tau_q$ is more sensitive to spatial correlation effects than the transport scattering time $\tau_t$ since $\tau_q$ is affected by the forward scattering process which is obviously more strongly influenced by spatial correlations. An immediate corollary of this finding is that the ratio $\tau_t/\tau_q$, which is very high ($\tau_t/\tau_q \gg 1$) in ultra high mobility modulation 
doped structures, would be much lower in the presence of spatial impurity correlations than in its absence. This is exactly the experimental observation, i.e., typically the theory based on completely uncorrelated random scattering by the remote charged impurities always gives calculated values of $\tau_t/\tau_q$ which are much larger than those found experimentally, and spatial correlations among the dopant ions could be (at least partially) a reason for this behavior.

The great advantage of the continuum model is its phenomenological nature where a single effective length parameter $r_0$ ($< (\pi n_d)^{-1/2}$) completely characterizes the spatial correlations
qualitatively and quantitatively. This great simplicity of the theory comes at the price that the continuum theory provides no explicit mechanism for calculating the phenomenological
 $r_0$-parameter from a microscopic model. In the next section, we consider a highly sophisticated (and highly numerically demanding) discrete lattice theory for dopant correlations which is completely microscopic and allows for an {\it ab initio} estimation of spatial correlations as a matter of principle (although as a matter of practice, all the growth parameters for modulation doped semiconductor structures are not known precisely enough for the discrete microscopic theory to lead to a purely theoretical quantitative evaluation of the spatial correlations).

\section{Discrete atomistic model of impurity correlations}

Molecular beam epitaxy (MBE) growth technique utilized in making ultrahigh mobility GaAs-AlGaAs semiconductor structures, being an atomistic growth process, entails a minimal spatial correlation among the deposited dopant atoms (e.g., the deposited dopant atoms must be separated by lattice distances from each other), but such a minimal spatial correlation is completely insufficient to describe the actual existing charged impurity correlations in the dopant layer. \cite{correlation,correlation_t} For example, we can put $r_0 = a_0$ (or even a few times $a_0$), where `$a_0$' is the typical unit cell size or lattice unit in GaAs or AlAs ($a_0 < 1$ nm), leading to $r_0/r_i \approx 0.1 - 0.3$, which (according to Figs.~\ref{fig3}--\ref{fig7}) should not have a very strong quantitative effect on the 2D mobility compared with the random scattering results shown in Fig.~\ref{fig1}. The atomistic physics of dopant impurity correlations must lie in some other physical mechanism for it to be relevant for our consideration of donor impurity correlation effects on the 2D mobility.

The key physical mechanism is provided by the fact that the dopant layer is not completely, but only partially, ionized, i.e., only a fraction `$f$' ($<1$) of the deposited dopant atoms in the modulation doping layer is ionized. This fractional ionization of the dopant layer very strongly enhances the possible spatial correlations among 
the charged or ionized donor impurities even if the spatial correlations 
among all the donor impurities (both charged and neutral) may be much weaker (as discussed above). If $f=1$, i.e., all the deposited donor atoms are somehow ionized in the modulation doping layer, then the impurity spatial correlations would indeed be rather small corresponding only to what the MBE growth process itself can impose (i.e., $r_0/r_i < 0.3$). Partial ionization of donors ($f<1$) allows donor electrons to hop around the donor atoms to find suitable donor sites in order to try to minimize the total energy of the whole collection of the charged donor atoms which interact with each other via the classical long-range Coulomb interaction whereas the neutral donors interact via weak short-range interaction. Such inter-donor hopping of electrons in order to achieve an apparent thermal equilibrium of the charged donor system is facilitated by the fact that high-quality MBE growth is a high-temperature process enabling inter-donor thermal activation of the electrons toward a local thermal equilibrium among the available donor impurities as enforced by the long-range Coulomb interaction.

To calculate the transport properties of a system with impurity correlations we need to calculate the pair correlation functions, $g_d(r)$ and $s_d(q)$. 
Once $g_d(r)$ and $s_d(q)$ are obtained, we can calculate scattering times ($\tau_{tq}$), mobility ($\mu$), and conductivity ($\sigma$) from the equations and formalism given in Sec. IV.
We  have calculated the pair correlation function $g_d({ r})$
for the ionized dopants using direct atomistic numerical Monte Carlo (MC) simulations \cite{correlation_mc,kodiyalam} on a system of charged dopants interacting via the classical Coulomb interaction on a 2D lattice. 
We consider a system of $N_i$ dopants obeying classical statistics, confined to an area $A$. The Hamiltonian, neglecting the kinetic energy, is given by 
\begin{equation}
H_I = \sum_{i<j} v(r_{ij}),
\end{equation}  
where $v(r_{ij}) = e^2/\kappa r_{ij}$ with $\kappa$ being the background dielectric constant and $r_{ij} = |{\bf r}_i -{\bf r}_j|$.
Then, the pair correlation function $g_d({\bf r})$ is defined by the equation
\begin{eqnarray}
g_d(r) = \frac{N_i(N_i-1)}{n_d^2 Q_{N_i}} \int \cdot \cdot \cdot  \int & & 
\exp \left [ -\beta \sum_{i<j}v(r_{ij}) \right ]   \nonumber \\
& &  d^2r_3 \cdot \cdot \cdot d^2r_{N_i},
\label{new_eq1}
\end{eqnarray}
where $\beta = 1/k_B T$, $n_d=N_i/A$ is the 2D dopant density, and
\begin{equation}
Q_{N_i}=\int \cdot \cdot \cdot \int \exp \left [ -\beta \sum_{i<j}v(r_{ij}) \right ] d^2r_1 \cdot \cdot \cdot d^2r_{N_i}.
\end{equation}
The MC simulations were carried out on a 2D square lattice of $N_i = N^2$ (with $N=32$ or 64 in most situation) where $N_i$ is number of ionized dopants (e.g., Si). These (infinitely heavy classical) charged particles were allowed to occupy points on a uniform fine grid that had $c^2$ ($c=42$ was used most often) points per unit area.
This discretization was used to enable the construction of a look-up table for the interaction potential between two arbitrary particles in the system. This obviously results in an upper bound for the wave vector values in the calculation. The potential used (and tabulated) is the Ewald summed ideal Coulomb interaction, which is the result of applying periodic boundary conditions to a parallelogram.

The simulations start with the particles in the ideal 2D Coulomb crystal (2D triangular lattice) configuration. We used the standard Metropolis MC algorithm to dynamically evolve the system with attempts to move particles serially one after another. The observed acceptance rate for our MC simulation is $\sim 0.65$.

\begin{figure}
	\centering
	\includegraphics[width=1.\columnwidth]{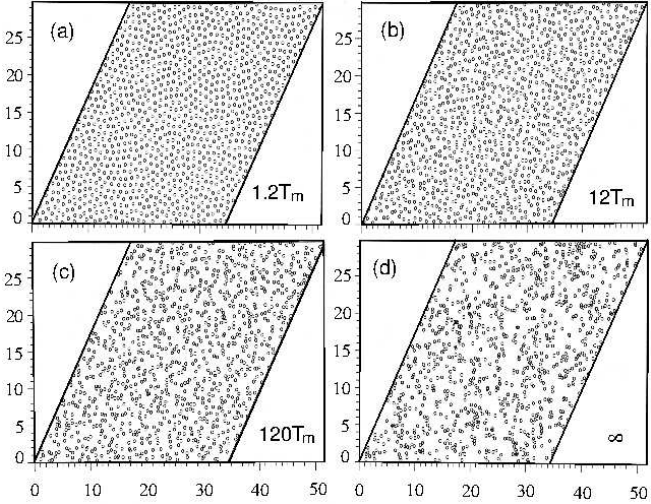}
	\caption{
The actual real space charged dopant distributions at the corresponding freezing temperatures: (a) 1.2 $T_m$, (b) 12 $T_m$, (c) 120 $T_m$, and (d) $\infty$.
The axes are labelled in units of $\sqrt{n_d}$.}
\label{fig9}
\end{figure}

\begin{figure}
	\centering
	\includegraphics[width=.9\columnwidth]{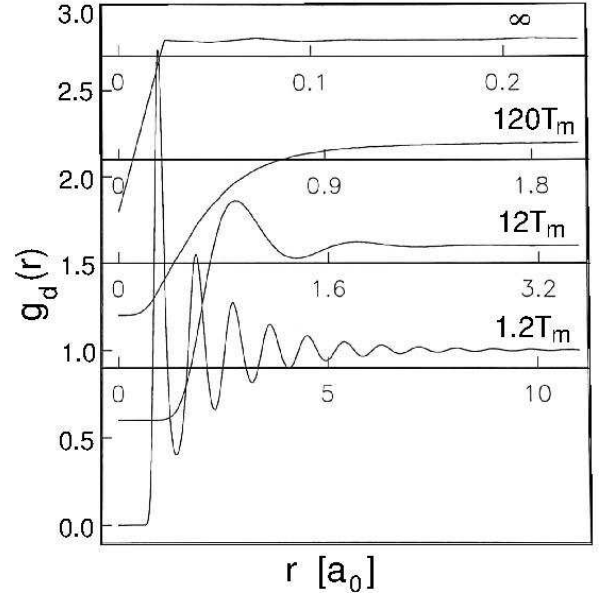}
	\caption{
The MC calculated pair correlation function $g_d({\bf r})$ as a function of the spatial separation $r$ between the ionized dopants for various values of the freezing temperature, 1.2 $T_m$, 12 $T_m$, 120 $T_m$, and $\infty$ (bottom to top). The results for higher temperatures are shifted upward for clarity. Here, $a_0 = r_0$. 
}
\label{fig10}
\end{figure}

\begin{figure*}
	\centering
	\includegraphics[width=1.6\columnwidth]{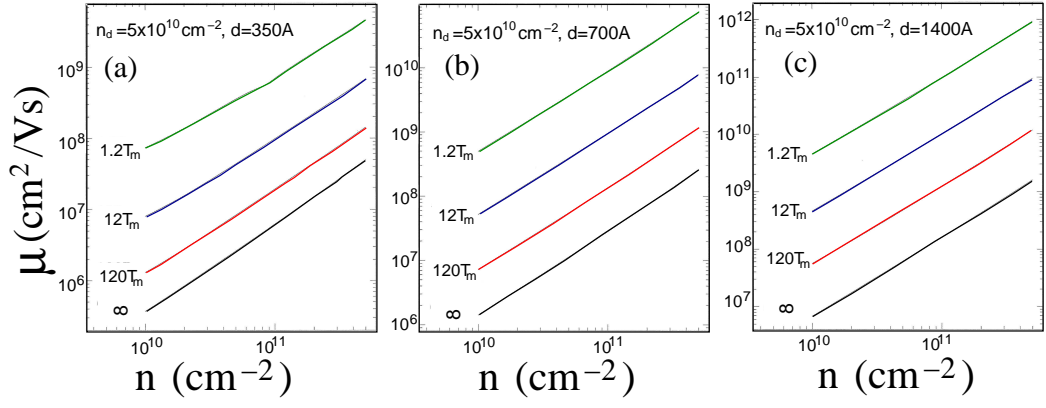}
	\caption{
The calculated mobility as a function of carrier density for a fixed value of ionized donor densities $n_d=5\times 10^{10}$ cm$^{-2}$  and for different set-back distances; (a) d=350 \AA, (b) d=700 \AA, and (c) d=1400 \AA. The calculation is done for four different temperatures,1.2 $T_m$, 12 $T_m$, 120 $T_m$, and $\infty$ (top to bottom). 
}
\label{fig11}
\end{figure*}

\begin{figure*}
	\centering
	\includegraphics[width=1.6\columnwidth]{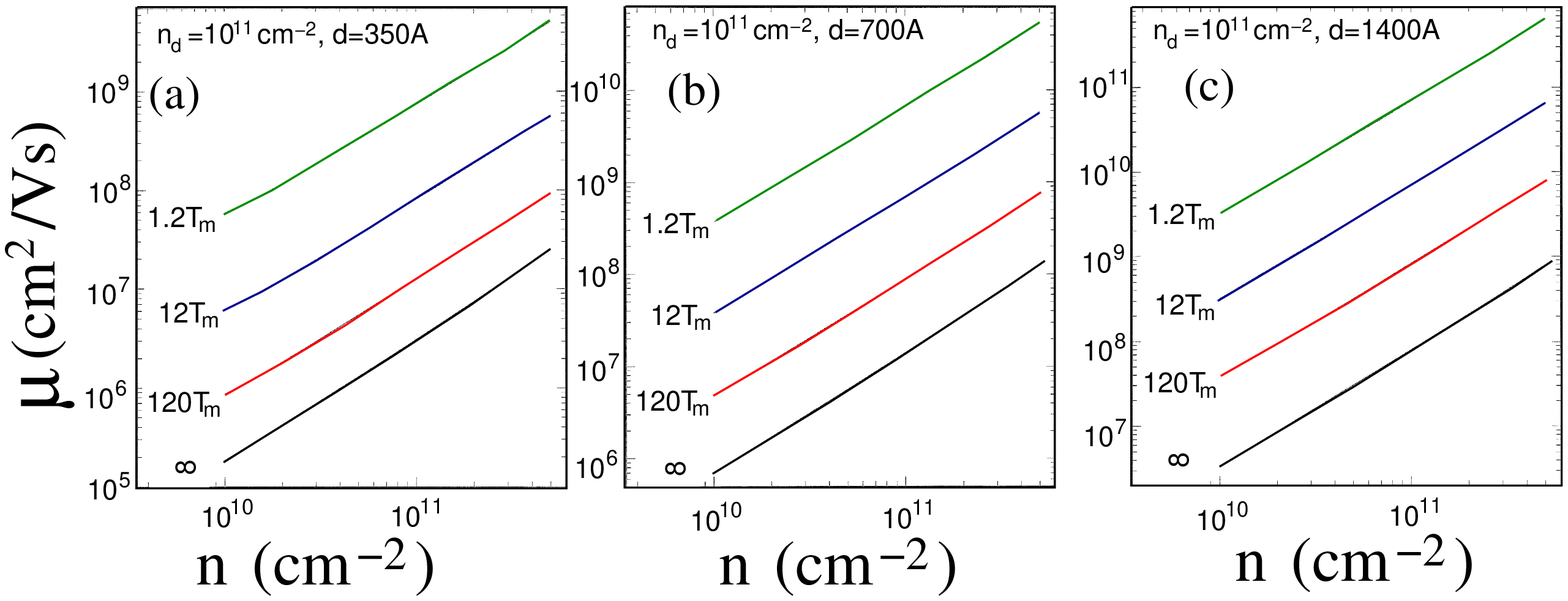}
	\caption{
The same as Fig.~\ref{fig11} except for $n_d=10^{11}$ cm$^{-2}$.
}
\label{fig12}
\end{figure*}

\begin{figure*}
	\centering
	\includegraphics[width=1.6\columnwidth]{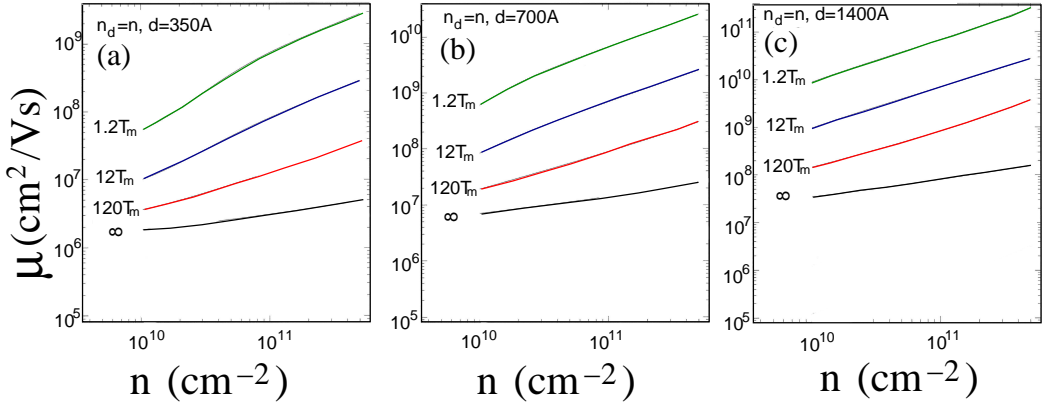}
	\caption{
The calculated mobility as a function of carrier density with $n_d= n$ and 
for different set-back distances: (a) $d=350$ \AA, (b) $d=700$ \AA, and (c) $d=1400$ \AA.
Each line indicates the different temperature as shown in the figure.
}
\label{fig13}
\end{figure*}

\begin{figure*}
	\centering
	\includegraphics[width=1.6\columnwidth]{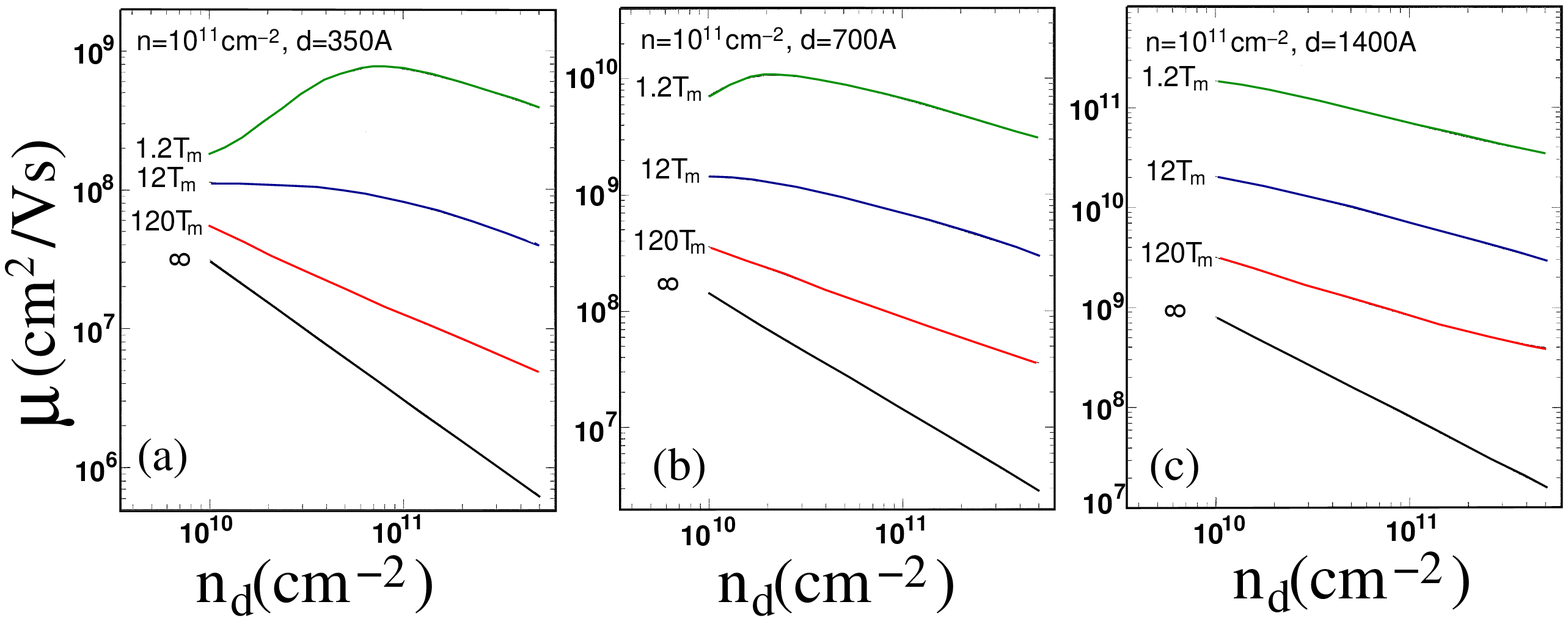}
	\caption{
The calculated mobility as a function of the ionized dopant density at a fixed carrier density $n=10^{11}$ cm$^{-2}$ and 
for different set-back distances: (a) $d=350$ \AA, (b) $d=700$ \AA, and (c) $d=1400$ \AA.
Each line indicates the different temperature as shown in the figure.
}
\label{fig14}
\end{figure*}

The single key variable (other than the ionized dopant density $n_d$) defining the converged results of the MC simulations (i.e., the calculated $g_d$ and $s_d$, for example) is the simulation ``temperature" ($T_0$), which is determined by the details of the MBE growth process (and not necessarily the growth temperature in the MBE chamber during the dopant impurity deposition in the modulation doping layer), and is the characteristic temperature at which the spatial correlations among the ionized dopants are frozen in. The basic idea is that, depending on the details of the prevailing growth conditions, electrons manage to hop around among the donor impurity sites so as to reach a global equilibrium at some temperature $T_0$ for the given ionized dopant density $n_d$ (which is smaller than the total dopant atom density allowing such an equilibrium configuration to be reached). Once this $T_0$-equilibrium is achieved, the system is frozen in and no further modifications in spatial correlations among the charged dopants can occur in the sample (unless, of course, the sample is heated to actual temperatures well above $T_0$ itself). Thus, $T_0$ is a ``nonequilibrium freezing or annealing temperature" where the spatial correlations set in and are frozen in. Unfortunately, $T_0$ itself is unknown (similar to $r_0$ in the continuum theory of Sec. IV) since it depends sensitively on the MBE growth conditions, but very crudely speaking $T_0$ should be of the order of the growth temperature (provided the growth temperature is not too low so that electrons can actually thermally hop around among the donor atoms to create thermal equilibrium at $T_0$) during the dopant incorporation.

The dimensionless quality characterizing the temperature $T_0$ is $T_0/T_m$ (similar to $r_0/r_i$ in the continuum model), where $k_B T_m = e^2(\pi n_d)^{1/2}$ is the melting temperature of 2D classical Coulomb crystal. Obviously $T_0/T_m > 1$ for the theory to be sensible because, otherwise (i.e., $T_0<T_m$), the ionized dopants will essentially form (if in equilibrium) a 2D crystalline solid leading to vanishing resistivity and infinite mobility (at least, as limited by remote dopant scattering). A more likely scenario for actual low-temperature ($<T_m$) growth is that the effective freezing temperature $T_0$ will be relatively high (and unknown) since the electrons are unlikely to be able to hop among the donor atoms to reach the global equilibrium periodic crystalline structure at the lower growth temperature forming instead a random glassy distribution of ionized donors which would correspond to an equilibrium structure for an elevated effective temperature $T_0$ ($>T_m$) and a nonequilibrium metastable structure at the actual growth temperature ($<T_m$). Our calculated inter-impurity correlations (i.e., $g_d$ and $s_d$) in the ionized donor spatial distribution will be characterized by $T_0/T_m$ with $T_0$ being the (generally unknown) nonequilibrium freezing temperature, which can be roughly taken as the MBE deposition temperature for the donor atoms in the modulation doping layer
(provided that the growth temperature is much higher than $T_m$ for the relevant ionized dopant density $n_d$ in the 2D doping layer, which is always the case).
More details on the MC simulation can be found in Ref. \onlinecite{kodiyalam}.

\begin{figure}
	\centering
	\includegraphics[width=.8\columnwidth]{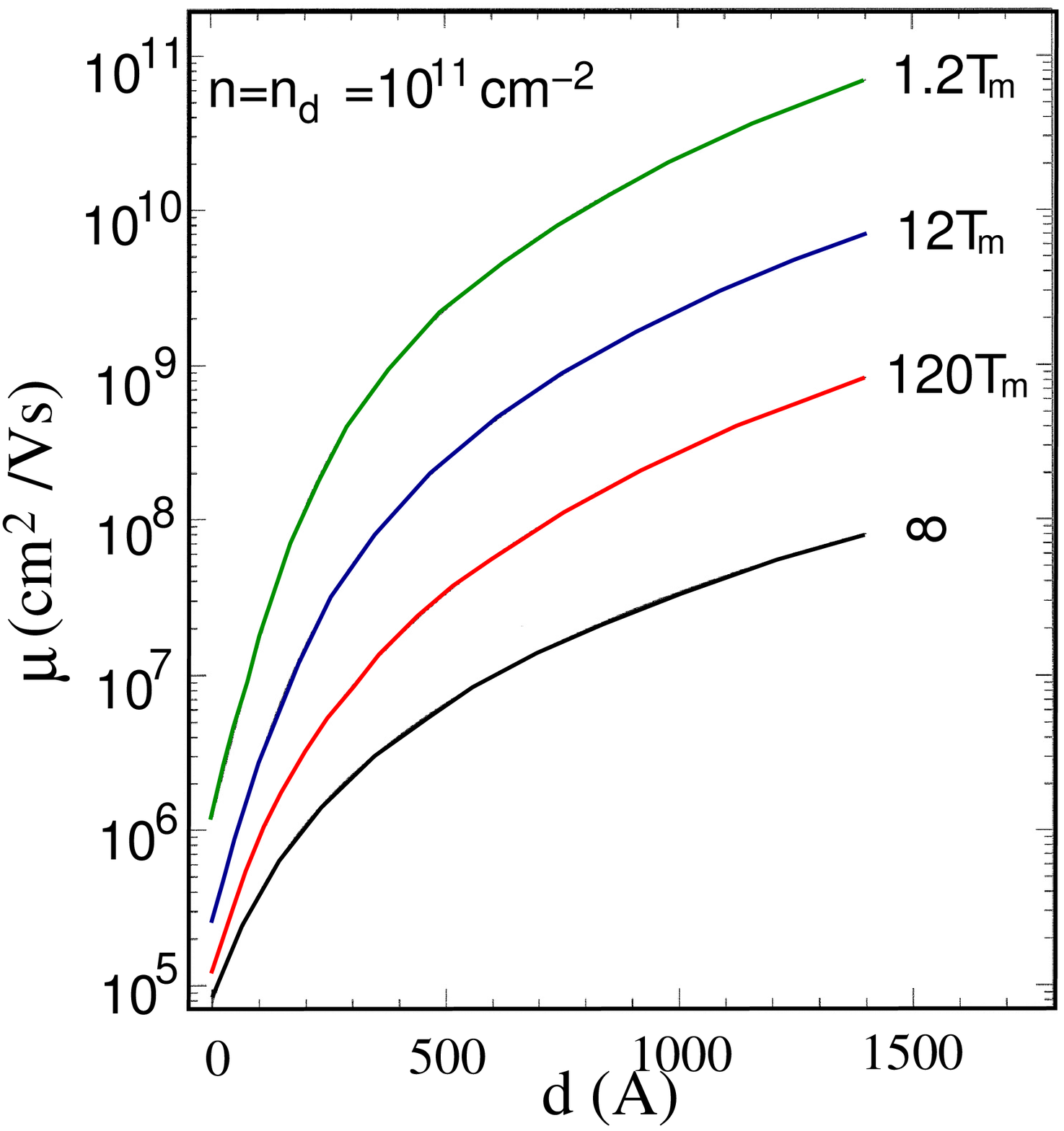}
	\caption{
The calculated mobility as a
function of the set-back distance (i.e., the separation between the dopant layer and the 2DEG) for $n=n_d=10^{11}$ cm$^{-2}$  and for different temperatures, 1.2 $T_m$, 12 $T_m$, 120 $T_m$, and $\infty$ (top to bottom).
}
\label{fig15}
\end{figure}

\begin{figure}
	\centering
	\includegraphics[width=.8\columnwidth]{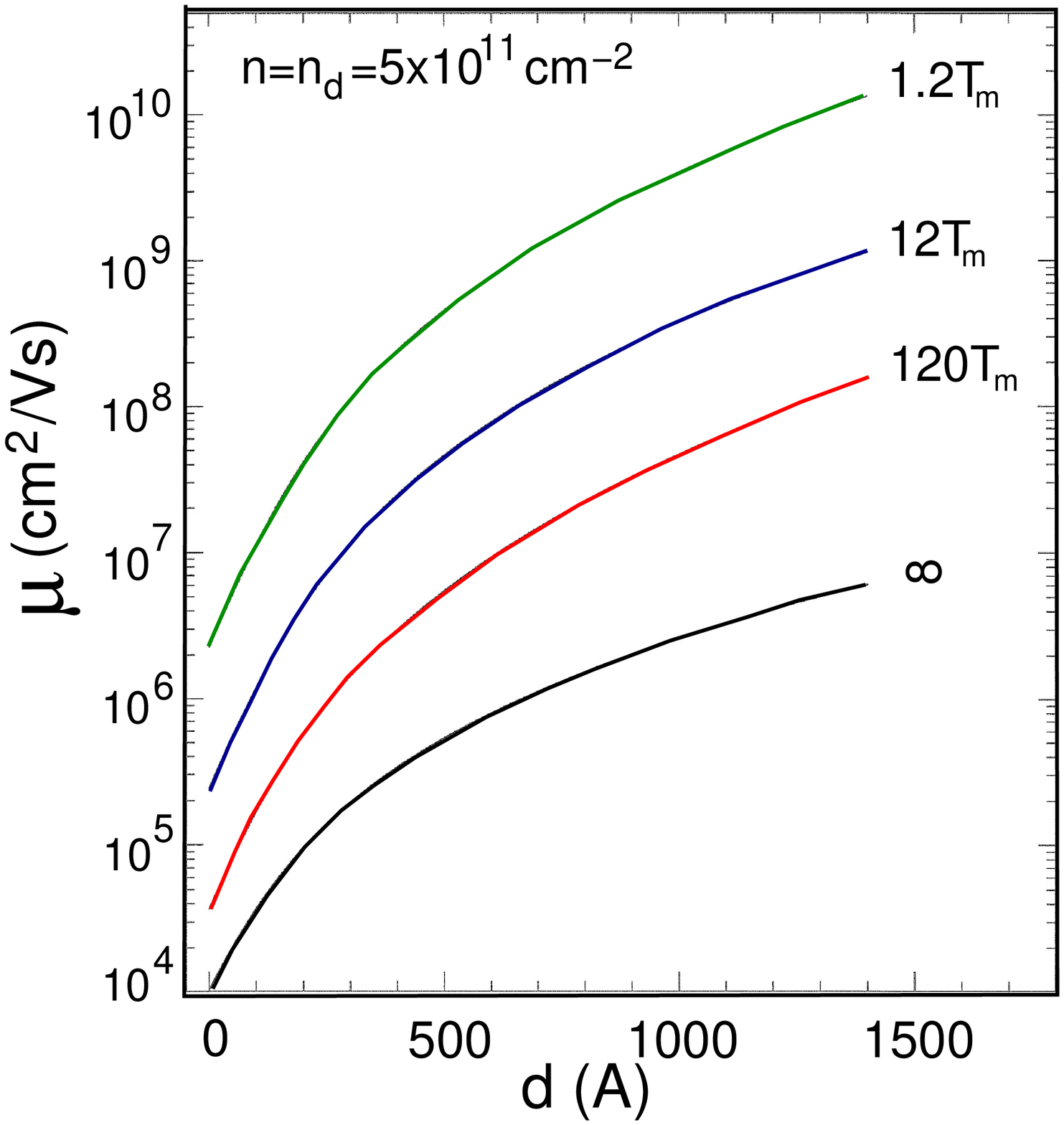}
	\caption{
The same as Fig.~\ref{fig15} but for $n=n_d=5\times 10^{11}$ cm$^{-2}$.
}
\label{fig16}
\end{figure}

To obtain the ionized dopant distribution from the MC simulation, we calculate the pair correlation function
$g_d({\bf r})$ and the associated structure factor $s_d({\bf q})$ for the dopant ions parametrized by the freezing temperature $T_0$ ($> T_m$) with $T_0 = \infty$ being the completely random uncorrelated impurity distribution with the correlations increasing strongly and monotonically as $T_0$ is lowered toward $T_m$ with $T_0 = T_m$ being the almost periodically perfectly correlated Coulomb crystalline phase of the dopant ions. 
In Fig.~\ref{fig9} we show the actual real space charged dopant distributions at the corresponding freezing temperatures as obtained from our detailed MC simulations.
In Fig.~\ref{fig10} we show the MC calculated pair correlation function $g_d({ r})$ as a function of the spatial separation $r$ between the ionized dopants for various values of the freezing temperature between $T_m$ and $\infty$. Clearly, the correlations are very strong for low temperatures (e.g., 1.2  $T_m$) and very weak for high temperatures (e.g., 120 $T_m$) with the infinite temperature situation showing no correlations whatsoever (i.e., the uncorrelated random impurities).

Once $g_d({r})$  is obtained from the MC calculations, the correlation structure factor, $s_d({q})$, is calculated as a two dimensional (fast) Fourier transform of $q_d(r)$.  
\begin{equation}
s(q) = \int_A g(r) e^{-i {\bf q \cdot r}} N_i d^2 r.
\end{equation}
Using the structure factor we can immediately calculate $\tau_t$, $\tau_q$, mobility ($\mu$) and conductivity ($\sigma$) from the equations and formalism already given in Sec. IV. 
In Figs.~\ref{fig11}--\ref{fig18} we show our representative numerical results for the spatially correlated  transport theory using the MC simulated values for the correlation structure factor $s_d({\bf q})$ characterized by the nonequilibrium freezing temperature.
We note that to calculate the quantum scattering time the MC simulation has been carried out in Ref.~[\onlinecite{correlation_mc}]. However, scattering time results were calculated as a function of charge state fluctuations. In this paper we provide the mobility results (and scattering times, $\tau_{t,q}$) in terms of all possible relevant parameters to experiment (i.e., $n$, $n_d$, $d$, and correlation parameter $T_0$).

In Figs.~\ref{fig11} and \ref{fig12}, we show the calculated mobility as a function of carrier density for two fixed values of ionized donor densities: $5\times 10^{10}$ cm$^{-2}$ (Fig.~\ref{fig11}) and $10^{11}$ cm$^{-2}$ (Fig.~\ref{fig12}). In each figure, the different panels give the results for different set-back distances (350, 700, 1400 \AA) as shown. Each figure provides results for four different sets of spatial correlations among the ionized dopants as characterized by four distinct  freezing temperatures (1.2$T_m$, $12T_m$, $120T_m$, and $\infty$). It is clear that spatial correlations could lead to huge enhancements of 2D mobilities compared with the uncorrelated random impurity scattering  results. 
Even a modest level of spatial correlations (e.g., $T_0=120$ $T_m$) produces an order of magnitude enhancement in the mobility (except for the smallest set back distance). 
The overall behavior of the mobility can be expressed by
\begin{equation}
\mu(n,d,n_d,T_0) \sim \frac{n^{3/2} d^3}{n_d^{1/2}} f(T_0/T_m),
\label{approx}
\end{equation}
where $f(T_0/T_m)$ is a function of the dimensionless parameter $T_0/T_m$. Apart from the function 
$f(T_0/T_m)$ the above approximation formula of the mobility describes very well the numerically calculated results in Figs.~\ref{fig11} and \ref{fig12}. The empirical form of the the function $f$ is given by $f(x) = (x+A)/(x-1)$, where $A\sim 10^3$ is the fitting constant.
By directly comparing these results with the results of the continuum model we find that $T_0/T_m \sim 10$, 100, $\infty$ approximately correspond to $(r_0/r_i)^2 \sim 0.99$, 0.9, 0, respectively.

In Fig.~\ref{fig13} we show the results as a function of carrier density making the standard $n= n_d$ assumption, again showing a large mobility enhancement due to spatial correlations. In this case ($n=n_d$) the mobility approximately increases with carrier density $n$ as $n^{1/2}$, as represented in Eq.~(\ref{approx}). 
In Fig.~\ref{fig14} we show the mobility as a function of the ionized dopant density at a fixed carrier density ($n=10^{11}$ cm$^{-2}$). 
We show in Figs.~\ref{fig15} and \ref{fig16} the calculated mobility as a
function of the set-back distance (i.e., the separation between the dopant layer and the 2DEG) for $n=n_d=10^{11}$ cm$^{-2}$ (Fig.~\ref{fig15}) and $n=n_d=5\times 10^{11}$ cm$^{-2}$ (Fig.~\ref{fig16}) for various spatially correlated impurity distributions. 

Finally in Figs.~\ref{fig17} and \ref{fig18} we show our calculated $\tau_t/\tau_q$ as a function of carrier density (Fig.~\ref{fig17}) and as a function of setback distance (Fig.~\ref{fig18}) including spatial correlation effects. 
The asymptotic behavior of the scattering time ratio 
is expressed by $\tau_t/\tau_q \sim n d^2$ for $k_Fd \gg 1$. This expression fits very well the numerically calculated results of Figs.~(\ref{fig17}) and (\ref{fig18}).
The most 
important qualitative message from Figs.~\ref{fig17} and \ref{fig18} is that even a modest amount of impurity correlations drastically reduces the ratio $\tau_t/\tau_q$ compared with its values for the completely random situation. This finding is in good qualitative agreement with all experimental measurements in the literature where $\tau_t/\tau_q$ is always found to be less than 100 in all high mobility systems independent of the values of carrier density and spacer thickness.

\begin{figure}
	\centering
	\includegraphics[width=1.\columnwidth]{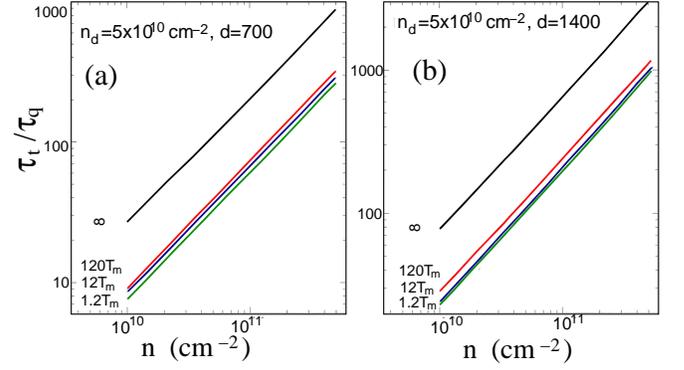}
	\caption{
The calculated scattering time ratio $\tau_t/\tau_q$ as a function of carrier density for $n_d=5\times 10^{10}$ cm$^{-2}$ and  for (a) $d=700$ \AA \; and (b) $d=1400$ \AA. The lines indicate the different temperatures: 1.2 $T_m$, 12 $T_m$, 120 $T_m$, and $\infty$ (bottom to top). 
}
\label{fig17}
\end{figure}

\begin{figure}
	\centering
	\includegraphics[width=.8\columnwidth]{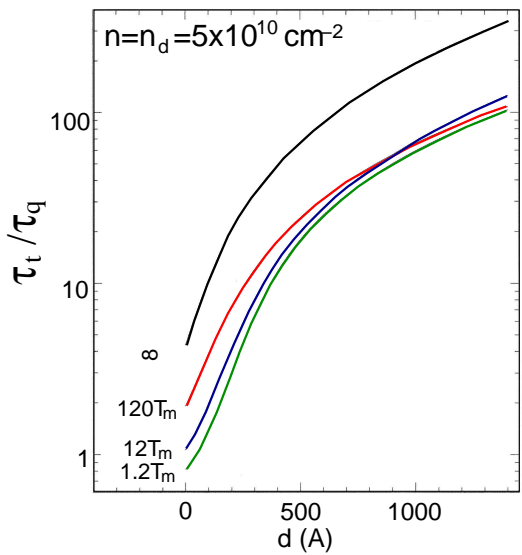}
	\caption{
The calculated $\tau_t/\tau_q$ as a function of set back distance for $n=n_d=5\times 10^{10}$ cm$^{-2}$ and  for different temperatures.
}
\label{fig18}
\end{figure}

Before concluding this section, it may be useful to discuss the expected range of $T_0$ in experimental samples. Of course, the precise value of $T_0$ depends on all the details of the sample growth and preparation conditions, and cannot therefore be reliably theoretically established. On the other hand, $T_m = e^2 (\pi n_d)^{1/2}$ can be expressed in degrees Kelvin as 
\begin{equation}
T_m(K) = e^2(\pi n_d)^{1/2} \approx 230 \tilde{n}_d,
\end{equation}
where $\tilde{n}_d = n_d/10^{12}$. Thus, $T_m = 230$ K (23 K) for $n_d=10^{12}$ ($10^{11}$) cm$^{-2}$. If we take $n_d \approx 5 \times 10^{11}$ cm$^{-2}$, we get $T_m \approx 115$ K, and a growth temperature of $500 - 700$ $^{\circ}$C then corresponds to $T_0 \agt 10 T_m$.
Given that the ultrahigh mobility 2D modulation doped structures typically have $n \agt 10^{11}$ cm$^{-2}$, we conclude that the reasonable range for the freezing temperature would be $T_0/T_m = 10 -100$. Thus, the actual mobility limited by remote scattering could easily be a factor of 10 or larger than the random scattering results given in Fig.~\ref{fig1}.
The precise estimate of the 2D mobility limited by the remote ionized dopant scattering  necessitates a precise knowledge of $T_0$, which is obviously unknown for the experimental samples.

\section{conclusion}

We have developed a theory for 2D transport in high-mobility modulation-doped structures assuming the resistive scattering to arise exclusively from remote ionized dopant scattering so that the calculated mobility is, by definition, the highest possible mobility in the system since all other resistive scattering mechanisms invariably present in the real structures are neglected. By carefully taking into account possible spatial correlations among the ionized dopant atoms in the modulation doping layer, 
we have tried to resolve the long-standing problem of why theoretically calculated highest possible mobility based on scattering from uncorrelated random disorder turns out to be always lower than the actual mobility values measured experimentally in ultrahigh mobility 2D modulation-doped structures. After establishing through a direct comparison with experiments that the Boltzmann transport theory along with the leading order Born approximation calculation for the scattering rate is an excellent quantitative approximation for high-mobility 2D systems, we show that spatial correlations among the ionized dopants typically enhance the calculated mobility by more than an order of magnitude compared with the purely uncorrelated random scattering model. Thus, for a 2D system with a density of $3\times 10^{11}$ cm$^{-2}$, the random  
scattering model predicts \cite{hwang2008} a mobility of $13\times 10^6$ cm$^2$/Vs for a 2D GaAs quantum well of width 300 \AA \; in a modulation doped structure with a set-back dopant layer separation of 800 \AA, to be compared with the maximum experimentally measured\cite{heiblum,bell} mobility of $36\times 10^6$ cm$^2$/Vs for the same sample parameters. Inclusion of reasonable spatial correlations among the ionized dopants enhances the mobility limited by remote dopant scattering to around $150-200\times 10^6$ cm$^2$/Vs indicating that remote Coulomb scattering is irrelevant for the transport properties of ultrahigh mobility modulation doped 
structures. 
This finding of the irrelevance of remote ionized dopant scattering in determining the 2D mobility in the best current modulation-doped samples has a number of implications beyond bringing experiment \cite{heiblum,bell} and theory \cite{hwang2008} together.  
It establishes that the mobility in the best current 2D modulation-doped structures is indeed limited by unintentional background charged impurities, as already demonstrated by the measured 
density-dependence of the mobility in the highest mobility 2D samples \cite{density} 
This leads to the possibility that the 2D mobility could be enhanced to as high as 100 million cm$^2$/Vs simply by purifying the background GaAs material for MBE growth since the maximum achievable mobility limited only by remote dopant scattering increases to $\sim 150$ million cm$^2$/Vs once the dopant impurity correlations are taken into account (in contrast to Ref.~\onlinecite{hwang2008} where the random dopant scattering puts a limit to the maximum mobility of around 13 million cm$^2$/Vs only).  The sensitive dependence of the 2D mobility on the spatial correlations among the remote ionized dopants also provides a possible explanation for the variation in the 2D mobility among various samples cut from the same 2D wafer for the highest mobility systems.\cite{pw}

An additional important result of our theory is that spatial correlations among the charged remote donors enhance $\tau_q$, the single-particle or the quantum scattering time, much more strongly than it enhances $\tau_t$, the mobility or the transport scattering time. This implies that the effective value of the ratio $\tau_t/\tau_q$ is much smaller ($\sim 10-100$) in the spatially correlated transport theory of modulation-doped structures than in the usual pure random scattering theory. We find that spatial correlations among the remote charged impurities always enhance $\tau_q$ more than it enhances $\tau_t$ since $\tau_q$ is much more sensitive to forward scattering which is strongly suppressed by correlation effects.

\section*{acknowledgement}
This work is supported by LPS-CMTC and ARO-IARPA.


\begin{thebibliography}{999}

\bibitem{nayakrmp} C. Nayak, S. H. Simon, A. Stern, M. Freedman, and S. Das Sarma,
Rev. Mod. Phys. {\bf 80}, 1083 (2008).

\bibitem{tsuiprl1982} 
K. von Klitzing, G. Dorda, and M. Pepper,
Phys. Rev. Lett. {\bf 45}, 494 (1980);
D. C. Tsui, H. L. St\"{o}rmer, and A. C. Gossard, Phys. Rev. Lett. {\bf 48}, 1559 (1982).

\bibitem{local} F. W. Van Keuls, X. L. Hu, H. W. Jiang, and A. J. Dahm,
Phys. Rev. B {\bf 56}, 1161 (1997); 
F. W. Van Keuls, H. Mathur, H. W. Jiang, and A. J. Dahm, 
Phys. Rev. B {\bf 56}, 13263 (1997).



\bibitem{mit}
M. J. Manfra, E. H. Hwang, S. Das Sarma, L. N. Pfeiffer, K. W. West, and A. M. Sergent, 
Phys. Rev. Lett. {\bf 99}, 236402 (2007);
S. Das Sarma and E. H. Hwang, Solid State Commun. {\bf 135}, 579 (2005);
S. Das Sarma and E. H. Hwang,
Phys. Rev. B {\bf 89}, 235423 (2014).


\bibitem{wigner} 
A. T. Hatke, Y. Liu, B. A. Magill, B. H. Moon, L. W. Engel, M. Shayegan, L. N. Pfeiffer, K. W. West	and K. W. Baldwin, Nature Commun. {\bf 5}, 5154 (2014); 
H. Zhu, Y. P. Chen, P. Jiang, L. W. Engel, D. C. Tsui, L. N. Pfeiffer, and K. W. West,
Phys. Rev. Lett. {\bf 105}, 126803 (2010).


\bibitem{anyon}
R. L. Willett, L. N. Pfeiffer, and K. W. West, PNAS {\bf 106}, 8853 (2008); 
I. P. Radu, J. B. Miller, C. M. Marcus, M. A. Kastner, L. N. Pfeiffer, and K. W. West, Science {\bf 320}, 899 (2008);
V. Venkatachalam, A. Yacoby, L. N. Pfeiffer, and K. W. West, Nature {\bf 469}, 185 (2011).




\bibitem{lillyprl} M. P. Lilly,  K. B. Cooper, J. P. Eisenstein, L. N. Pfeiffer, and K. W. West, Phys. Rev. Lett. {\bf 82}, 394 (1999).


\bibitem{micro} R. G. Mani, J. M. Smet, K. von Klitzing, V. Narayanamurti, W. B. Johnson, and V. Umansky, Nature {\bf 420}, 645 (2002); M. A. Zudov, Phys. Rev. B {\bf 69}, 041304 (2004);
I. A. Dmitriev, A. D. Mirlin, D. G. Polyakov, and M. A. Zudov, Rev. Mod. Phys. {\bf 84}, 1709 (2012).

\bibitem{drag} T. J. Gramila, J. P. Eisenstein, A. H. MacDonald, L. N. Pfeiffer, and
K. W. West, Phys. Rev. Lett. {\bf 66}, 1216 (1991); A. G. Rojo, J. Phys. Condens. Matter {\bf 11}, R31 (1999); E. H. Hwang, S. Das Sarma, V. Braude, and A. Stern, Phys. Rev.
Lett. {\bf 90}, 086801 (2003).

\bibitem{spon} S. Q. Murphy, J. P. Eisenstein, G. S. Boebinger, L. N. Pfeiffer, and K. W. West, Phys. Rev. Lett. {\bf }72, 728 (1994);
I. B. Spielman, J. P. Eisenstein, L. N. Pfeiffer, and  K. W. West, Phys. Rev. Lett.  {\bf 84}, 5808 (2000).

\bibitem{canted}
A. Fukuda, A. Sawada, S. Kozumi, D. Terasawa, Y. Shimoda, Z. F. Ezawa, N. Kumada, and Y. Hirayama, Phys. Rev. B {\bf 73}, 165304 (2006);
L. Brey, E. Demler, and S. Das Sarma, Phys. Rev. Lett. {\bf 83}, 168 (1999);
S. Das Sarma, S. Sachdev, and L. Zheng, Phys. Rev. Lett. {\bf 79}, 917 (1997);
Phys. Rev. B {\bf 58}, 4672 (1998);
L. Zheng, R. J. Radtke, S. Das Sarma, Phys. Rev. Lett. {\bf 78}, 2453 (1997).

\bibitem{exciton}
Y. W. Suen, L. W. Engel, M. B. Santos, M. Shayegan, and D. C. Tsui,
Phys. Rev. Lett. {\bf 68}, 1379 (1992); 
J. P. Eisenstein, G. S. Boebinger, L. N. Pfeiffer, K. W. West, and Song He,
Phys. Rev. Lett. {\bf 68}, 1383 (1992).


\bibitem{willettprl1987}
R. Willett, J. P. Eisenstein, H. L. St\"{o}rmer, D. C. Tsui, A. C. Gossard, and J. H. English, Phys. Rev. Lett. {\bf 59}, 1776 (1987);
C. R. Dean, B. A. Piot, P. Hayden, S. Das Sarma, G. Gervais, L. N. Pfeiffer, and K. W. West, Phys. Rev. Lett. {\bf 100}, 146803 (2008); {\bf 101}, 186806 (2008); 
H. C. Choi, W. Kang, S. Das Sarma, L. N. Pfeiffer, and K. W. West,  Phys. Rev. B {\bf 77}, 081301(R)
  (2008).
  


\bibitem{heiblum} V. Umansky, M. Heiblum, Y. Levinson, J. Smet, J. N\"{u}bler, M. Dolev, J. Cryst. Growth {\bf 311}, 1658 (2009);
M. Dolev, M. Heiblum, V. Umansky, A. Stern, and D.  Mahalu,
  Nature {\bf 452}, 829 (2008);  V. Umansky, R. de Picciotto, and M. Heiblum, Apple. Phys. Lett. {\bf 71}, 683 (1997).


\bibitem{bell}
J. P. Eisenstein, K. B. Cooper, L. N. Pfeiffer, and K. W. West, Phys. Rev. Lett. {\bf 88}, 076801 (2002);
W. Pan, J. S. Xia, H. L. Stormer, D. C. Tsui, C. Vicente, E. D. Adams, N. S. Sullivan, L. N. Pfeiffer, K. W. Baldwin, and K. W. West, Phys. Rev. B {\bf 77}, 075307 (2008);  
W. E. Chickering, J. P. Eisenstein, L. N. Pfeiffer, and K. W. West,
Phys. Rev. B 87, 075302 (2013); 
N. Deng, A. Kumar, M. J. Manfra, L. N. Pfeiffer, K. W. West, and G. A. Cs\'{a}thy, 
Phys. Rev. Lett. 108, 086803 (2012); 
Yanhua Dai, R. R. Du, L. N. Pfeiffer, and K. W. West,
Phys. Rev. Lett. {\bf 105}, 246802 (2010);
M. Manfra, arXiv:1309.2717;  L. Pfeiffer and K.W West, Physica E {\bf 20}, 57 (2003).



\bibitem{stormer} H. L. St\"{o}rmer, R. Dingle, A. C. Gossard, and
  Wiegmann, Inst. Conf. Ser. London {\bf 43}, 557 (1978).


\bibitem{quality} S. Das Sarma and  E. H. Hwang,
Phys. Rev. B {\bf 90}, 035425 (2014).

\bibitem{hwang2008} E. H. Hwang and S. Das Sarma,  Phys. Rev. B {\bf 77}, 235437 (2008).

\bibitem{stern1983}  F. Stern, Appl. Phys. Lett. {\bf 43}, 974 (1983).

\bibitem{pfeiffer1989} L. Pfeiffer, K. W. West, H. L. Stormer and K. W. Baldwin,
Appl. Phys. Lett. {\bf 55}, 1888 (1989).


\bibitem{correlation} 
A. F. J. Levi, S. L. McCall and P. M. Platzman, Appl. Phys. Lett. {\bf 54}, 940 (1989);
A. L. Efros, F. G. Pikus, and G. G. Samsonidze, Phys. Rev. B {\bf 41}, 8295 (1990);
T. Kawamura and S. Das Sarma, Solid State Commun. {\bf 100}, 411 (1996); 
V. M. Mikheev, Phys. Solid State {\bf 49}, 1856 (2007);
Qiuzi Li, E. H. Hwang, E. Rossi, and S. Das Sarma,
Phys. Rev. Lett. {\bf 107}, 156601 (2011).

\bibitem{correlation_t} E. Buks, M. Heiblum, Y. Levinson, and H. Shtrikman, Semicond. Sci. Technol. {\bf 9}, 2031 (1994); E. Buks, M. Heiblum, and Hadas Shtrikman,
Phys. Rev. B {\bf 49}, 14790(R) (1994).

\bibitem{correlation_mc} R. Shikler, M. Heiblum, and V. Umansky, Phys. Rev. B {\bf 55}, 15 427 (1997).

\bibitem{kodiyalam} S Das Sarma and S Kodiyalam, Semicond. Sci. Technol. {\bf 13}, A59 (1998).


\bibitem{andormp} T. Ando, A. B. Fowler, and F. Stern,
  Rev. Mod. Phys. {\bf 54},  437 (1982); S. Das Sarma, S. Adam, E. H. Hwang, and E. Rossi, Rev. Mod. Phys. {\bf 83}, 407 (2011).

\bibitem{lilly} 
B. E. Kane, L. N. Pfeiffer, K. W. West, and C. K. Harnett, Appl. Phys. Lett. {\bf 63}, 2132 (1993);
M. P. Lilly, J. L. Reno, J. A. Simmons, I. B. Spielman, J.
P. Eisenstein, L. N. Pfeiffer, K. W. West, E. H. Hwang,
and S. Das Sarma, Phys. Rev. Lett. {\bf 90}, 056806 (2003);
S. Das Sarma, M. P. Lilly, E. H. Hwang, L. N. Pfeiffer,
K. W. West, and J. L. Reno, Phys. Rev. Lett. {\bf 94}, 136401 (2005);


\bibitem{density} S. Das Sarma and E. H. Hwang, Phys. Rev. B {\bf 88}, 035439 (2013).


\bibitem{fluctuation} B. I. Shklovskii and A. L. Efros, Zh. Eksp. Teor. Fiz. {\bf 61}, 816 (1971) [Sov. Phys. JETP {\bf 34}, 435 (1972)]; {\it Electronic Properties of Doped
Semiconductors} (Springer-Verlag, Berlin, 1984); 
S. Das Sarma, E. H. Hwang, and Q. Li, Phys. Rev. B {\bf 88}, 155310 (2013).


\bibitem{dassarmaPRB1985} S. Das Sarma and F. Stern, Phys. Rev. B {\bf 32}, 8442 (1985).


\bibitem{pw}  L. N. Pfeiffer and K. W. West, unpublished.


\end{thebibliography}
\end{document}